\documentclass[useAMS,usenatbib]{mnras} 
\usepackage{graphicx}
\usepackage{amssymb, amsmath}
\usepackage{times}
\usepackage{color}
\usepackage{todonotes}
\usepackage{lscape}
\usepackage{booktabs}
\usepackage[section]{placeins}
\usepackage{breqn}
\usepackage{setspace}
\usepackage{adjustbox}

\usepackage{lineno}

\usepackage{arydshln}
\usepackage{url}

\definecolor{carrotorange}{rgb}{0.93, 0.57, 0.13}

\newcommand{\gtsima}{$\; \buildrel > \over \sim \;$}
\newcommand{\ltsima}{$\; \buildrel < \over \sim \;$}
\newcommand{\simgt}{\lower.7ex\hbox{\gtsima}}
\newcommand{\simlt}{\lower.7ex\hbox{\ltsima}}

\newcommand{\orcid}[1]{\href{https://orcid.org/#1}{\includegraphics[width=0.7em]{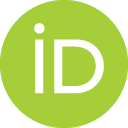}}}

\def\vaststrut{\vrule width0pt height2.5em}
\def\medstrut{\vrule width0pt height1.8em}
\def\f{\frac}
\def\mpcoh{\,h^{-1}{\rm Mpc}}
\def\kpcoh{\,h^{-1}{\rm kpc}}
\def\msolaroh{\,h^{-1}M_\odot}
\def\citejap#1{\citeauthor{#1}\ \citeyear{#1}}
\def\m@th{\mathsurround=0pt }
\def\eqalign#1{\null\,\vcenter{\openup1\jot \m@th
 \ialign{\strut\hfil$\displaystyle{##}$&$\displaystyle{{}##}$\hfil
 \crcr#1\crcr}}\,}

\begin{document}


\title[Satellite galaxies and RSD]{Using GAMA to probe the impact of small-scale galaxy physics on nonlinear redshift-space distortions}

\author[Alam et. al.] {\parbox{\textwidth}{
    Shadab Alam\orcid{0000-0002-3757-6359}$^{1}$ \thanks{ salam@roe.ac.uk},
    John A. Peacock\orcid{0000-0002-1168-8299}$^{1}$\thanks{jap@roe.ac.uk},
    Daniel J. Farrow$^{2}$,
    J. Loveday\orcid{0000-0001-5290-8940}$^{3}$,
    and A. M. Hopkins$^{4}$ 
    } \vspace*{4pt} \\ 
\vspace{-1.5mm} $^{1}$Institute for Astronomy, University of Edinburgh, Royal Observatory, Blackford Hill, Edinburgh, EH9 3HJ , UK\\
\vspace{-1.5mm} $^{2}$Max-Planck-Institut f{\"u}r extraterrestrische Physik, Giessenbachstrasse 1, 85748 Garching, Germany\\
\vspace{-1.5mm}  $^{3}$Astronomy Centre, University of Sussex, Falmer, Brighton BN1 9QH, UK\\ 
\vspace{-1.5mm} $^{4}$Australian Astronomical Optics, Macquarie University, 105 Delhi Rd, North Ryde, NSW 2113, Australia
}

\date{\today}
\pagerange{\pageref{firstpage}--\pageref{lastpage}}   \pubyear{2018}
\maketitle
\label{firstpage}

\begin{abstract}
    We present improved modelling of the redshift-space distortions of galaxy clustering that arise from peculiar velocities. We create mock galaxy catalogues in the framework of the halo model, using data from the Bolshoi project. These mock galaxy populations are inserted into the haloes with additional degrees of freedom that govern spatial and kinematical biases of the galaxy populations relative to the dark matter. We explore this generalised halo model with an MCMC algorithm, comparing the predictions to data from the Galaxy And Mass Assembly (GAMA) survey, and thus derive one of the first constraints on the detailed kinematic degrees of freedom for satellite galaxies within haloes.
    With this approach, the distortions of the redshift-space galaxy autocorrelations can be accounted for down to spatial separations close to 10 kpc, opening the prospect of improved RSD measurements of the perturbation growth rate by the inclusion of data from nonlinear scales. 
\end{abstract}


\section{Introduction}
\label{sec:intro}

It is fascinating that Cosmic Microwave Background (CMB) measurements reveal a simple picture of the early universe, with a thermal spectrum \citep{1996ApJ...473..576F} and Gaussian temperature fluctuations
\citep{2011ApJS..192...18K,2018arXiv180706209P,2013ApJ...779...86S,2013JCAP...10..060S}\footnote{\url{https://lambda.gsfc.nasa.gov/product/}}. The $\Lambda$CDM model successfully matches these extremely precise measurements, with General relativity \citep{Einstein1915} at its core -- although the inferred dominant constituents of dark energy \citep{Riess1998,Perlmutter1999} and dark matter \citep{Zwicky1937, Kahn1959, Rubin1970} remain obscure from a fundamental physics point of view. The lack of theoretical understanding of the dark sector, plus unresolved questions regarding the early universe, motivates modified theories of gravity (e.g. \citejap{2010AnPhy.325.1479J},
\citejap{2012PhR...513....1C}); but no such alternative theory without dark matter and dark energy has been as successful as $\Lambda$CDM. 

The $\Lambda$CDM framework provides well specified initial condition of the universe, which yield a simple prediction of structure formation. The initial density fluctuations observed in the CMB will grow under gravity leading to the formation of structure at late times (e.g. \citejap{Comer1994}). These initial density fluctuations remain small for most of the cosmic history and hence can be solved by linear perturbation theory \citep{Mukhanov1992,Liddle1993, Durrer1994, Ma1995, Bruni1994, Kopeikin2001,Bernardeau2002, Lagos2016}. The $n$-point clustering of the late time dark matter distribution has very specific properties and hence is sensitive to the physical quantities and dynamical equations of the universe \citep{1970ApJ...162..815P,Eis2005, Bassett2010,Coil2013}.  To a good approximation, galaxies form at peaks of the dark matter density distribution, thus tracing the $n$-point functions of the dark matter distribution up to a multiplicative constant called galaxy bias \citep{Bardeen1986,Cole1989}. 

Modern cosmological surveys can create three dimensional maps of these density fluctuations by measuring redshifts for large numbers of galaxies.
Past spectroscopic galaxy surveys (2dFGRS: \citejap{Colless2003}; 6dFGS: \citejap{Jones2009}) have measured $\sim 10^5$ redshifts; more recent surveys (SDSS-III: \citejap{Eisenstein2011};  WiggleZ: \citejap{WiggleZ}; DEEP2: \citejap{Deep2013}; VIPERS: \citejap{Garilli2014};  GAMA: \citejap{gama2015}; SDSS-IV: \citejap{2016AJ....151...44D}) have been of the same size or up to $\sim 10^6$ redshifts; near-future surveys (PFS: \citejap{2014PASJ...66R...1T}; 4MOST: \citejap{2019Msngr.175....3D}; DESI: \citejap{2016arXiv161100036D}) will measure $\sim10^7$ galaxy spectra. Such datasets provide in particular exquisitely precise measurements of the $n$-point statistics of the galaxy distribution. The observed two-point correlation function (2PCF) of galaxies in such a survey is anisotropic due to redshift space distortions (RSD). The galaxies have peculiar velocities of their own due to their gravitational interaction with the overall dark matter field, and thus the observed redshift is a combination of the uniform Hubble expansion and the peculiar velocity along the line-of-sight (los). This gives a radial distortion in the galaxy 2PCF along the los compared to the orthogonal direction. The theoretical existence of RSD was discussed in \citet{Peebles1980} and \citet{Kaiser87} developed the linear theory formalism for RSD in Fourier space, which was translated to configuration space by \citet{Hamilton92}.  Several authors have tried to go beyond linear theory to model RSD in recent surveys \citep[e.g.][]{2008PhRvD..78h3519M,Carlson12,2014JCAP...05..003O,2016JCAP...12..007V}, but it has been hard to push to small scales of 1~Mpc or less using such an approach: we are then far from the regime of perturbation theory for the dark-matter dynamics, and on these scales the assumption of modelling galaxy formation with a single bias parameter is inevitably too simplistic. Another approach to modelling more non-linear scales is to use a fully non-linear solution for the mass distribution through dark matter only N-body simulations, followed by modelling galaxy formation physics via an empirical galaxy-halo connection. One important attempt to use such an approach on SDSS-III data is described in \cite{Reid14}, who obtained $2.5\%$ precision on the growth rate using autocorrelation data down to $1\mpcoh$. Recently there have been attempts to develop emulators that can efficiently span the cosmological parameters as well as parameters of the galaxy-halo connection to predict the clustering, for example \citet{2018arXiv180405867Z}. The main focus of these studies has been to extract the measurement of the growth rate from redshift-space galaxy clustering, treating galaxy physics as a nuisance to be marginalized over.

Such modelling neglects the data below about 1 Mpc; and yet we have very precise measurements of the galaxy 2PCF at small scales, and it is interesting to ask what extra insights about both galaxy formation and cosmology can be obtained by moving beyond the arena of quasilinear large-scale structure. This is the regime we explore in the present paper, covering galaxy clustering from kpc to Mpc scales. Specifically, we analyze clustering from the GAMA survey, covering scales as small as $0.01\mpcoh$. Although other surveys cover larger volumes, GAMA is unique in being essentially complete spectroscopically even to such small scales, whereas its competitors suffer a systematic loss of very close pairs of galaxies. In this paper we model small scale clustering using the halo occupation distribution \citep[HOD;][]{Benson2000,Seljak2000,Peacock2000,White2001,Berlind2002,Cooray2002} in conjunction with dark-matter halo catalogues from numerical simulations. By using high-resolution data that are complete to low halo masses, we are able to synthesize a much deeper galaxy samples than used in past studies, which helps us address some of the details concerning the distinct populations of central and satellite galaxies. Finally, the mocks created as the result of this study should be useful for exploring questions about galaxy groups, colours and assembly bias in future work. 

The paper is organized as follows. We first introduce the GAMA data and how we create several magnitude limited samples in section \ref{sec:data}. In section \ref{sec:model} we describe the details of our modelling methodologies, followed by our measurements from the GAMA data in section \ref{sec:measurement}. The section \ref{sec:analysis} summarises our analysis methodologies with results being presented in section \ref{sec:result}. We present a summary and discussion in section \ref{sec:summary}.

We assumes a flat $\Lambda$CDM cosmology with $\Omega_m=0.27$, $\Omega_b=0.0469$, $h=0.7$, $n_s=0.95$ and $\sigma_8=0.82$. Our assumed cosmology is motivated by the fiducial cosmology assumed in the $N$-body simulation \citep[Bolshoi;][]{2011ApJ...740..102K} 
that we employ in our HOD models. 

\section{Data}
\label{sec:data}

We use data from the GAMA survey 
described in \cite{2015MNRAS.452.2087L} and \cite{2018MNRAS.474.3875B}. GAMA is a flux limited spectroscopic survey of approximately 300,000 galaxies (215,260 made public in DR3), selected from SDSS imaging \citep{2009ApJS..182..543A} with input catalogue defined in \cite{2010MNRAS.404...86B}, covering a total sky area of $230$ deg$^2$. It has a redshift completeness of 98\% down to $r$-band Petrosian magnitude of 19.8. The tiling strategy is explained in \cite{2010PASA...27...76R} and the spectroscopic pipeline is described in \cite{2013MNRAS.430.2047H}.  We are using three GAMA equatorial regions, namely G09, G12 and G15, centred on 9h, 12h and 14.5h in right ascension, each consisting of $5\times12\,{\rm deg}^2$. We first define a k-corrected and evolution corrected absolute magnitude in order to create three magnitude limited samples with $M_r<-21$, $M_r<-20$ and $M_r<-19$. The following section describes the details of our sample selection.

\subsection{Galaxy subsamples}
We use the k-corrected $r$-band Petrosian magnitude of galaxies to account for the fact that observed magnitudes of galaxies at different redshifts will probe different parts of the galaxy spectral energy distribution. The k-corrections were derived using template spectra and galaxy magnitudes in the 5 SDSS photometric bands, as described in \cite{Blanton2003} and \cite{2012MNRAS.420.1239L, 2015MNRAS.451.1540L}.

\begin{figure}
    \begin{center}
    \includegraphics[width=0.48\textwidth]{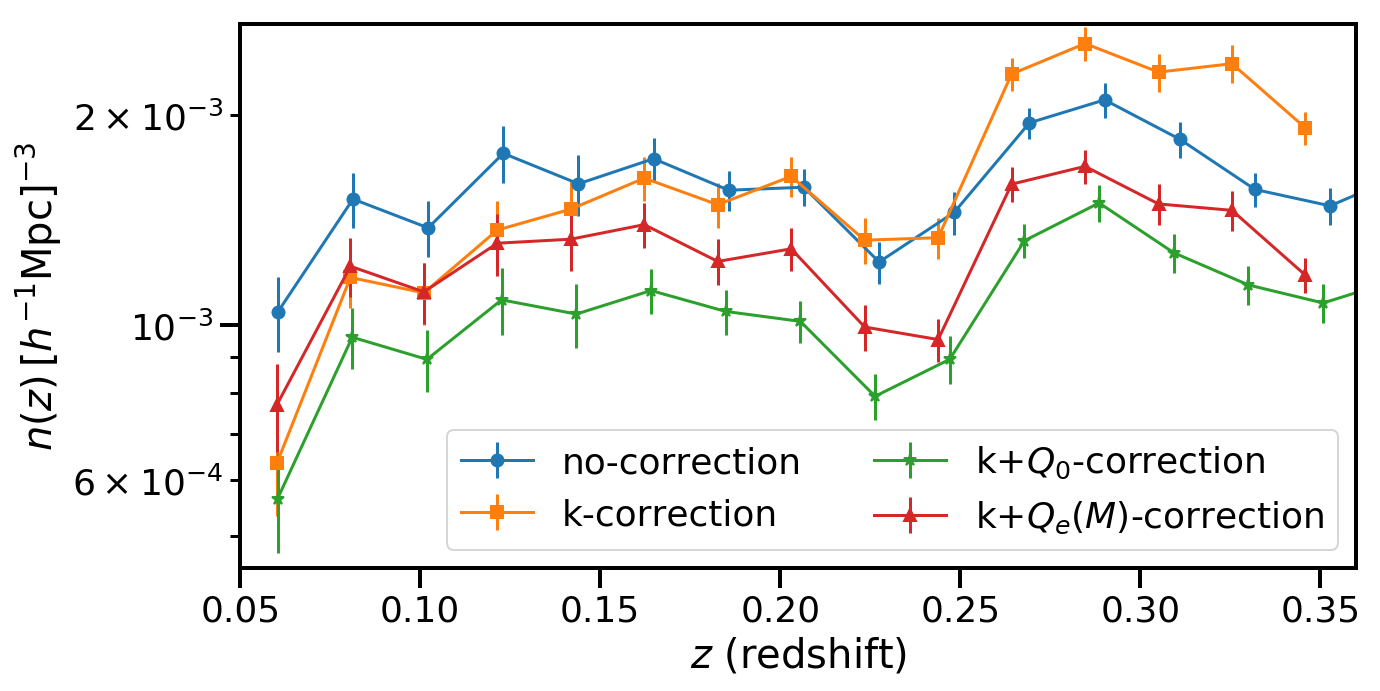}
    \caption{Number density of $M_r<-21$ galaxies as a function of redshift with various corrections to magnitude, showing jackknife errors. The circle, square, star and triangle symbols represent respectively density with no-correction, with only k-correction, with $k+Q_0$ correction and with $k+Q_e(M)$ correction. Here, $Q_0$ is the evolution correction with reference at $z=0$ while $Q_e(M)$ is the magnitude dependent evolution correction with reference redshift at $z=0.1$.}
    \label{fig:nz}
    \end{center}
\end{figure}

The observed magnitude is also affected by luminosity evolution. We have taken this into account by applying a redshift-dependent correction to the magnitudes so that
the galaxy sample is consistent in comoving density over the entire range of redshift. We have used two different ways to correct luminosity evolution and found consistent results for our sample. We first used the luminosity evolution correction $E(z)=-Q_0 (z-z_{\rm ref})$ , where $Q_0=0.97$ and $z_{\rm ref}=0$ derived in \citet{2014MNRAS.445.2125M}. We have also derived a new luminosity-dependent luminosity evolution correction $E(z,M)=Q_e(M)(z-z_{\rm ref})$. where $M$ is the magnitude with k-correction but without any evolution correction, $Q_e(M)$ is given by a sigmoid function which corresponds to approximately $Q_0=1.0$ for the brighter sample given by the previous correction. We find that the magnitude dependence of $Q_e(M)$ is best described by a sigmoid function
In this correction we used $z_{\rm ref}=0.1$ to be consistent with the reference redshift used in the k-correction. The  expression giving $Q_e(M)$ is as follows:
\begin{equation}
Q_e(M)=A/(1.0+e^{-W\times (M+19)}) + Q_{\rm shift},
\end{equation}
where $A=-5.8$, $W=4.5$ and $Q_{\rm shift}=-1.09$; these values were obtained by fitting a linear model of luminosity evolution with redshift in magnitude bins. The r band absolute magnitude is given by:
\begin{equation}
    M_r -5 \log_{\rm 10}h = r_{\rm petro} -5 \log_{\rm 10}\left( \frac{d_L(z)}{h^{-1}Mpc}\right) -25 -k(z)- E(z)
\end{equation}
where $r_{\rm petro}$ is the r band petrosian flux, $d_L$ is the luminosity distance, $k(z)$ is the k-correction and $E(z)$ is the luminsoity evolution. We finally define a magnitude-limited sample of galaxies in the redshift range 0.05 to 0.36  and $M_r<-21$, where $M_r$ is $r$-band absolute magnitude after applying the k-correction and the magnitude dependent evolution correction. This gives us 23309 galaxies in the three GAMA regions G09, G12 and G15. Figure \ref{fig:nz} shows the number density as function of redshift for galaxies with $M_r<-21$. The different points show magnitude limited samples defined using the various correction terms. We note that the two different methods for correcting the luminosity evolution give the same result, apart from a small vertical shift which is due to the difference in the reference redshift assumed for the two cases. As stated above, we adopt a reference redshift of 0.1. We also create two other magnitude limited samples $M_r<-20$ and $M_r<-19$ for extending our analysis to deeper samples but smaller volumes.

\section{Modelling non-linear scales}
\label{sec:model}

The linear theory of large-scale structure (LSS) successfully describes many aspects of the distribution of matter in the universe. But linear theory becomes progressively less applicable on small scales and at late times when the structure undergoes nonlinear growth. The remarkable success of linear theory can be extended to gain further information by solving for higher order terms in perturbation theory and using data on quasi-linear scales \citep{Socco2004,2010PhRvD..82f3522T,2011JCAP...08..012O,ReiWhi11,2012MNRAS.427.2537C,2012JCAP...11..009V}. But the only fully reliable way to model non-linear scales is to solve for the exact dark-matter dynamics via N-body simulations. Several methods exist for performing pure dark matter N-body simulations in efficient ways and these have been shown to give quite robust predictions \citep{2011ApJ...740..102K,2016MNRAS.457.4340K}. But such simulations can only tell us about dark matter haloes and their distribution in the universe. In order to model the observed galaxy distribution we must understand how galaxies form and evolve within the scaffolding of dark matter. The direct approach to this problem is to use a full hydrodynamical simulation \citep[e.g.][]{2010MNRAS.402.1536S,2014Natur.509..177V,2016MNRAS.463.3948D,2017arXiv170609899T}, where all the components of the universe are accounted for and galaxies form through a variety of physical processes. Such simulations are currently rather limited in nature as it is impractical to simulate a large cosmological volume while retaining sufficient resolution to capture the details of galaxy formation. Therefore the omitted small-scale processes have to be re-inserted in an approximate fashion using a range of `subgrid' recipes. This leads to a range of results from various groups working on such simulations, and the calculations cannot currently be considered to have converged at the level we would need in order to analyse real data \citet{2018MNRAS.480.3962C}. 

A common alternative approach is therefore to work with large volume dark matter only simulations, populating them with galaxies based on empirical models. Two such models widely used in LSS analyses are subhalo abundance matching (SHAM; see e.g. \citejap{2004MNRAS.353..189V}; \citejap{2006ApJ...647..201C}) and the Halo Occupation Distribution (HOD), discussed below. The former is more demanding computationally, as it requires sufficient resolution to identify the substructure within dark-matter haloes.
In this paper we therefore use the HOD approach to populate galaxies within a dark matter simulation and compare the observables calculated from such mock catalogues with data. In the following section we describe the details of different components of our model.

\subsection{N-body simulation and halo catalogue}

We are using the publicly available Bolshoi 
simulation \cite{2011ApJ...740..102K} through the CosmoSim database \footnote{\url{https://www.cosmosim.org/cms/simulations/bolshoi/}}. Bolshoi is a dark matter only N-body simulation run using the Adaptive-Refinement-Tree (ART) code \citet{1997ApJS..111...73K}. The simulation assumes a flat $\Lambda$CDM cosmology with $\Omega_m=0.27$, $\Omega_b=0.0469$, $h=0.7$, $n_s=0.95$ and $\sigma_8=0.82$. It is in a periodic box of side length 250$\mpcoh$  and $2048^3$ particles. The ART code used for Bolshoi is designed to preserve the physical resolution to $\sim7\kpcoh$ for $z=0-8$.

A halo catalogue using the ROCKSTAR\footnote{\url{https://bitbucket.org/gfcstanford/rockstar}} halo finder \citet{behroozi13} was constructed using the snapshot at an effective redshift of $z = 0.1$ for Bolshoi. ROCKSTAR starts with a friends-of-friends group catalogue and analyses particles in full phase space (i.e. position and velocity) in order to define halo properties and robustly identify substructures. We use particle data from MDPL1\footnote{\url{https://www.cosmosim.org/cms/simulations/mdpl/}} at $z=0$.

\subsection{Galaxy-halo connection}
It is widely accepted that galaxies form in the densest regions of universe. As cosmological density perturbations become nonlinear, they generate dark matter haloes, which provide the appropriate environment for baryonic processes to form stars and hence galaxies. Such processes are more effective in high-mass haloes, and as a result of this bias the clustering of galaxies shows a higher amplitude than the clustering of matter itself. At large linear scales, this yields galaxy $n$-point clustering statistics that are a simple multiple of the underlying matter statistics. But as we look at smaller non-linear scales, gravitational collapse leads to non-linear evolution of the matter power spectrum. Thus on these non-linear scales the halo or galaxy power spectrum in general displays a scale-dependent bias in the clustering signal. In order to model this non-linear bias of galaxies we choose to populate haloes with galaxies, using the Halo Occupation Distribution approach
\citep[HOD:][]{Benson2000,Seljak2000,Peacock2000,White2001,Berlind2002,Cooray2002}. This places two distinct kinds of galaxies into haloes: a single central galaxy that dominates the halo, plus a number of distinct satellites.

We use the HOD method to model the observed clustering of GAMA galaxies within the Bolshoi dark matter haloes. The HOD model
used in this paper was proposed in \cite{White2011}, and it assumes a specific functional form with a small number of free parameters for the occupation probability of a halo with a central and satellite galaxies:
\begin{eqnarray}
\left\langle N_{\rm cen} \right\rangle_M &= \frac{1}{2}  \mathrm{erfc}\left( \frac{\ln (M_{\rm cut}/M)}{\sqrt{2}\sigma_{\rm M}}\right) \, , \nonumber \\
\left\langle N_{\rm sat}\right\rangle_M &=  N_{\rm cen}(M) \left(  \frac{M-\kappa M_{\rm cut}}{M_1}\right)^\alpha \,
\label{eqn:HOD}\, ,
\end{eqnarray}
where $\left\langle N_{\rm cen}\right\rangle_M$ gives the occupation probability of central galaxies in a halo of given mass $M$ and average number of satellite galaxies is given by $\left\langle N_{\rm sat}\right\rangle_M$ .

The occupation probability of central galaxies involves an error function with two parameters $M_{\rm cut}$ and $\sigma_M$. This is motivated by the fact that we are trying to model a magnitude limited sample. So our galaxy selection is a step function in magnitude. If we assume that there exists a monotonic relationship between the absolute magnitude of a galaxy and its host halo mass then we should be able to model these galaxies with a similar step function in halo mass where the parameter $M_{\rm cut}$ sets the location of this halo mass limit for the corresponding magnitude limit. But such a relation between halo mass and light from the central galaxy will inevitably have some intrinsic scatter that depends on the exact formation history of the halo (and which dominates over any measuring error, although both would combine into a single empirical dispersion). Therefore one should allow the occupation probability to `turn on' over some width which is given by $\sigma_M$. The assigned number of central galaxies is one or zero, chosen as a binomial quantity according to the occupation probability.

The number of satellite galaxies in a halo is sampled from a Poisson distribution with mean given by $\left\langle N_{\rm sat}\right\rangle_M$. 
The parameter $\kappa$ specifies the minimum halo mass to host a satellite galaxy in units of $M_c$. The parameter combination $M_1+\kappa M_c$ gives the halo mass at which the halo on average will have a single satellite galaxy. The more massive haloes are expected to host numerous satellite galaxies and the model assumes that this number scales as a power law of mass with index $\alpha$. This parametrization is motivated from hydrodynamical simulation and semi-analytic models of galaxy formation \citep{zheng2005}.

Once we know how many central and satellite galaxies a given halo should host we need to be able to assign their position with respect to the halo. We place central galaxies at the centre of mass of the halo and satellite galaxies are distributed with an NFW radial profile in a spherically symmetric manner:
\begin{eqnarray}
    \rho(r) &=&  \f{\rho_0}{\f{r}{R_s} \left(1+\f{r}{R_s}\right)^2} \label{eqn:nfw}\\
    \rho_0&=&\f{M_{\rm vir}}{4 \pi R_s^3 \left[\ln(1+c)-\f{c}{1+c}\right]}\, .
    \label{eqn:rho0}
\end{eqnarray}
The above equations define the NFW profile \citep{1996ApJ...462..563N}, whose parameters $R_s$, $c$ and $M_{\rm vir}$ are scale radius, concentration and virial mass of the halo respectively; these are measured by fitting the NFW profile to individual halo density profiles as part of the ROCKSTAR analysis. By definition, $R_s\equiv r_{\rm vir}/c$, where the virial radius $r_{\rm vir}$ is the outer edge of the halo.
We populate satellites within $r_{\rm vir}$ using an NFW distribution with an additional free parameter $f_{c}$, which is a multiplicative factor to the concentration, allowing the satellite distribution to deviate from the dark matter halo distribution. This $f_c$ parameter defines a new scale radius for the satellites as $R'_s = R_s/f_c$. This is physically motivated by looking at hydrodynamical simulations, which indicate that the stellar and gas distribution could be different from the dark matter distribution \citet{Zhu_2016}. But in any case it is safer not to assume that satellite
galaxies must follow the dark matter exactly, even though this is a common assumption in HOD analyses.

\subsection{Redshift space distortions and galaxy velocities}

Spectroscopic surveys such as GAMA measure the spectral energy distribution (SED) of their target galaxies, yielding a precise redshift for each galaxy from the relative shift in the location of spectral features. In order to measure three-dimensional clustering of these galaxies one needs a radial distance, which is inferred from the redshift assuming the distance--redshift relation of an isotropic fiducial cosmology. But in addition to the redshift from cosmological expansion, the observed redshift also contains imprints from galaxy dynamics and its environment. The observed redshift can be written as sum of the Hubble recession velocity, galaxy peculiar velocity, local gravitational potential and other relativistic effects \citep{Cappi1995,Yoo2014,2017MNRAS.470.2822A,2017MNRAS.471.2345Z}:
\begin{equation}
    z_{\rm obs} = H(z)r/c + v_{\rm pec}/c + z_g + \cdots\, ,
    \label{eqn:redshift}
\end{equation}
where $z_{\rm obs}$ is the observed redshift, r is the true line-of-sight distance, $H(z)$ is the Hubble parameter,  $v_{\rm pec}$ is the radial peculiar velocity of the galaxy, $c$ is the speed of light and $z_g$ is the gravitational redshift. The above expression is valid for distances r where a linear Hubble relation is a good approximation; in general it is better to say that $v_{\rm pec}$ adds an offset $(1+z)v_{\rm pec}/H(z)$ to the inferred comoving distance of the galaxy. The gravitational redshift is an order of magnitude smaller than the peculiar velocity \citep{2017MNRAS.471.2345Z}. Therefore we will ignore any impact of gravitation redshift and higher order terms on the observed redshift in this paper. The peculiar velocity systematically biases the line-of-sight distance estimated from the observed redshift, breaking the spherical symmetry of large scale structure traced by galaxies. This makes the two-point galaxy clustering anisotropic and is known as redshift space distortion (RSD).

\begin{figure}
    \centering
    \includegraphics[width=0.48\textwidth]{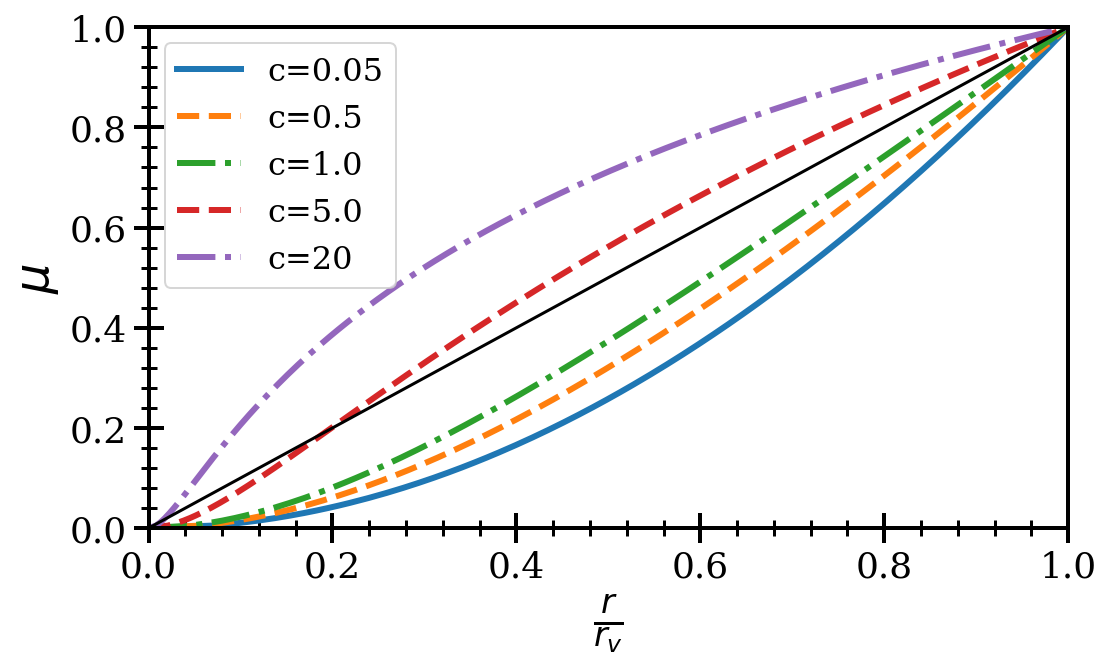}
    \caption{Relationship between halo radius in units of virial radius ($r/r_v$) with $\mu$ indicating fractional halo mass enclosed within the distance. The different lines show different values of concentration parameter as indicated in the legend. The black solid line represents a linear relation to guide the eye. } 
    \label{fig:r-mu}
\end{figure}

Therefore in order to model the observed galaxy survey in redshift space one must be able to model the galaxy velocities accurately. We study the structure of dark matter halo velocities in the MPL1 N-body simulation in order to quantify the uncertainty and motivate our modelling choices regarding galaxy velocities. We first define a measure of distance from halo centre as the fraction of halo mass enclosed ($\mu$). This allows us to average haloes with different concentration and virial radius in a consistent manner within a range of halo mass. One can easily transform $\mu$ to distance from halo centre assuming the NFW halo density profile by solving the following equation:
\begin{equation}
\ln(1+cx)-\frac{cx}{1+cx}=\mu \left[ \ln(1+c)-\frac{c}{1+c} \right] \, ,
    \label{eq: Men-rad}
\end{equation}
where $c$ is the halo concentration parameter and $x=r/r_v$ with $r_v$ being the virial radius of the dark matter halo. We show a set of solutions to the above equation in Figure~\ref{fig:r-mu} for various concentration parameters. This illustrates how different values of $r/r_v$ map to our measure of distance $\mu$ used in the next two subsections.

\subsubsection{Central galaxy velocity}

The first question we ask is to how to assign the velocity of central galaxies. For each halo in the simulation we measure the core velocity $v_{\rm core}$, which is defined as the mean velocity of 10\% of the particles in the halo around the halo centre with a minimum of 100 particles. We use the core velocity as the velocity of the central galaxy. In order to estimate the impact of this choice on the velocity of the central galaxy, we look at the mean velocity of the halo particle as a function of fractional halo mass enclosed ($\mu$). 

\begin{figure}
    \centering
    \includegraphics[width=0.48\textwidth]{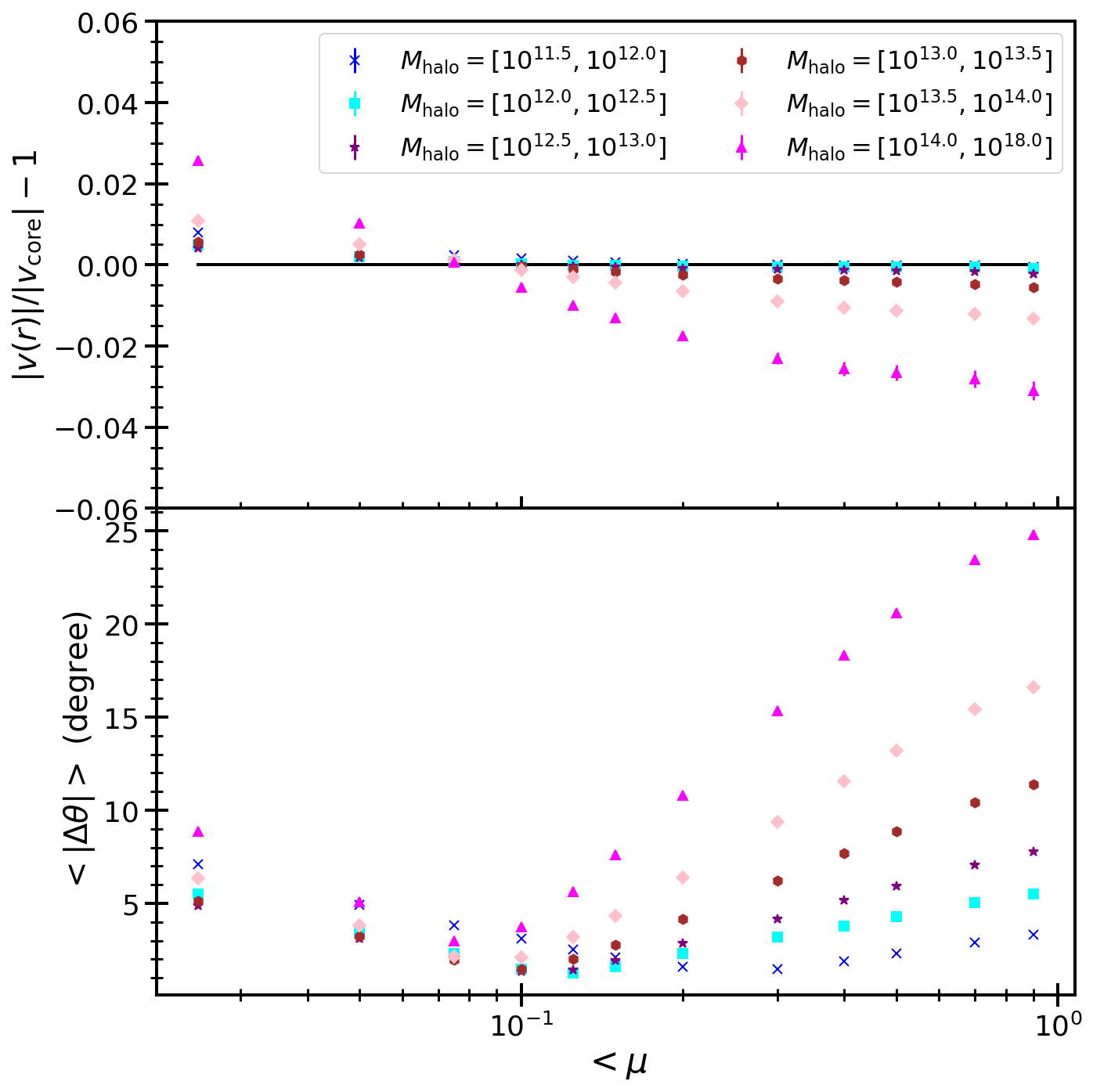}
    \caption{Magnitude and angle of the offset between core velocity as a function of halo mass enclosed ($<\mu$), relative to $\mu=0.1$. The top panel shows that the amplitude of core velocity is robust against fraction of halo mass enclosed within 1\% except in the most massive haloes where this increases to 2-3\%. The bottom panel shows that the mean halo velocity has a slight misalignment from the centre of the halo to the outskirts at the level of $5 \deg$ which increase to larger values outside 50\% of mass enclosed. 
     } 
    \label{fig:vcen-mu}
\end{figure}

The top panel of Figure \ref{fig:vcen-mu} shows how the core velocity varies in magnitude for different fractions of halo mass enclosed, as measured in the MDPL1 simulation. The peculiar velocity of the central part of the halo is uncertain at the 1\% level depending on what scale we measure, with this difference rising to 2--3\% in the most massive haloes. Such an impact is negligible for our analysis, but we will include a 2\% additional systematic uncertainty in our measurement due to the uncertainty in the choice of central galaxy velocity. The second panel of Figure \ref{fig:vcen-mu} also shows how the direction of the velocities changes as we include more particles. This indicates a change of direction of less than 5$^\circ$ for up to 50\% of the halo particles, which increases to larger values in the outer part of the halo; the amount of misalignment between the outer part of the halo and the inner part is proportional to the mass of the halo. Therefore, we expect our constraints to be largely independent of the choices we made in the definition of core velocity which is assigned to the central galaxies.

\subsubsection{Satellite galaxy velocities}

The next question is how to assign the velocities of the satellite galaxies. In our default model we assign satellite velocities from a normal distribution $\mathcal{N}(v_{\rm core},\gamma_{\rm IHV} \sigma_v)$ where $\sigma_v$ is the halo velocity dispersion up to virial radius and $\gamma_{\rm IHV}$ is a free parameter which allows satellite galaxy velocity distribution to be different from that of the dark matter in the halo. Note that we only use the velocity dispersion of satellites along the line of sight, with a Gaussian distribution. In principle this function will also depend on the relative position within the main halo with respect to the los. As a result a combination of tangential and radial velocity dispersions will contribute to the satellite velocity distribution. We discuss this point in more detail below.

\begin{figure}
    \centering
    \includegraphics[width=0.48\textwidth]{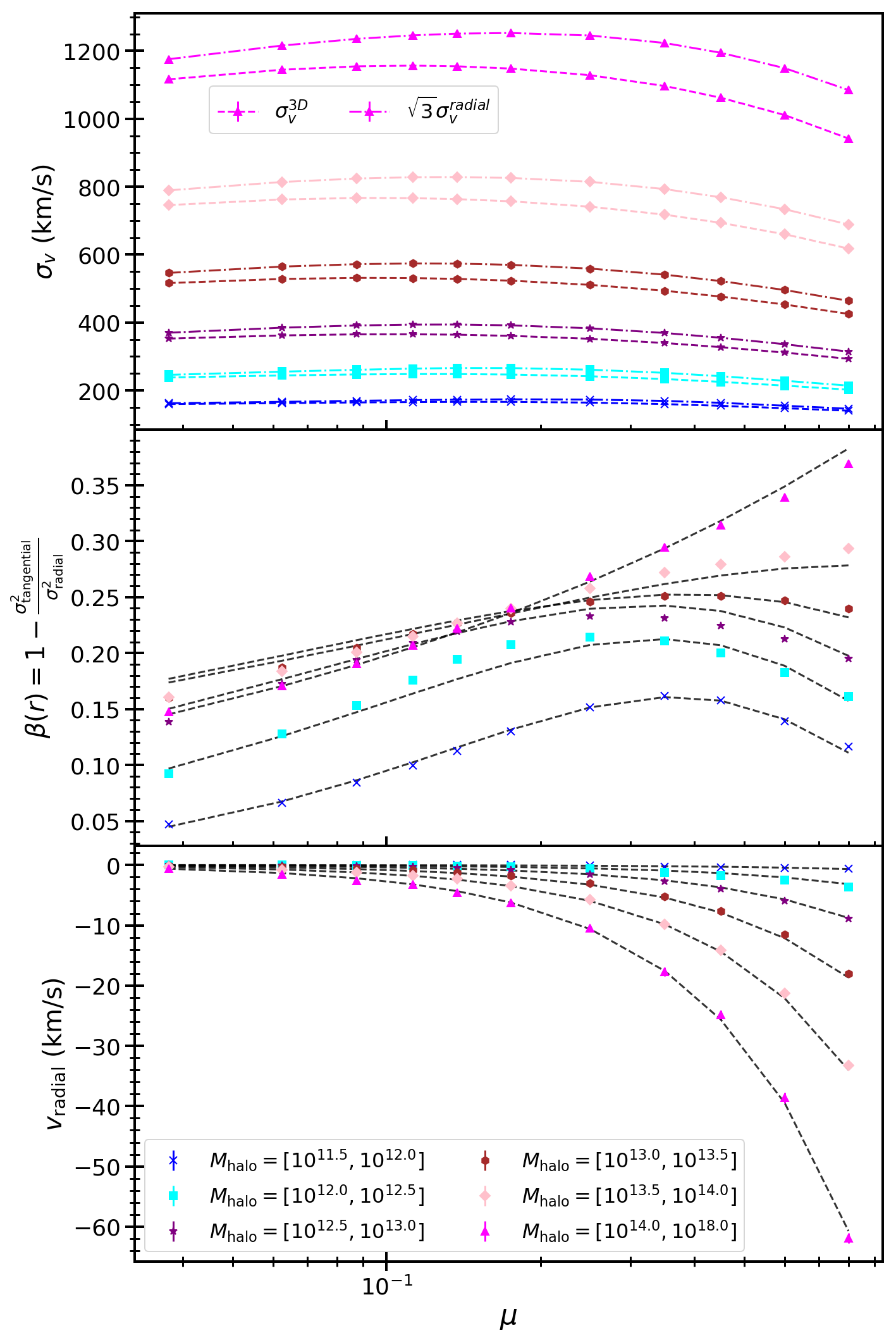}
    \caption{The velocity statistics of the dark matter halo particles as a function of halo mass. The top, middle and bottom panels show the velocity dispersion, velocity anisotropy and radial velocity respectively, as a function of mass enclosed ($\mu$). The different coloured markers represent haloes in different ranges of mass as indicated in the legend. In the top panel the points connected via dashed lines show the three dimensional velocity dispersion and the points connected via dotted-dashed lines refer to radial velocity dispersion. In all three panels, the points shows measurements from N-body simulations. The dashed black lines in the middle and bottom panel are from the empirical models proposed in this paper. 
     } 
    \label{fig:vsat-mu}
\end{figure}

Figure~\ref{fig:vsat-mu} shows the velocity statistics of particles in dark matter haloes measured from N-body simulations as a function of the enclosed mass fraction ($\mu$). The different coloured markers represent different halo mass bins as indicated in the legend. The top panel shows three-dimensional velocity dispersion with dashed lines and radial velocity dispersion with dashed-dotted lines. Note that the velocity dispersion strongly increases with halo mass but is constant with respect to distance from the halo centre. The radial velocity dispersion is greater than or equal to the three-dimensional velocity dispersion, indicating the presence of velocity anisotropy together with a net infalling motion. 

The middle panel of Figure~\ref{fig:vsat-mu} shows the velocity anisotropy measured from the N-body simulation with markers. \cite{2001MNRAS.321..155L} suggested using $\beta(s)=s^2/(s^2+s_a^2)$ as the parametric form to fit velocity anisotropy where $s$ is the distance from halo centre and $s_a$ is a free parameter called anisotropy radius. This functional form fails to describe the velocity anisotropy measured in our simulations. \cite{2008ApJ...682..835Z} studied velocity anisotropy by solving the Jeans equation with the assumption of a spherical dark matter halo in dynamical equilibrium with the NFW density and a power-law phase space density profile. They predict a functional form for the velocity anisotropy parameter $\beta$, given in their equation 14. We found that this functional form does not provide a good fit to the measured velocity anisotropy in N-body simulations either. This is probably due to a failure of some of the assumptions in the analytic model. Therefore, we propose the following empirical relation for the velocity anisotropy as the function of halo mass and the mass enclosed parameter $\mu$:
\begin{align}
\beta(\mu,{\rm M_{h}})& = A_{\beta}({\rm M_{h}}) \times \mu^{B_{\beta}({\rm M_{h}})} \times e^{-C_{\beta}({\rm M_{h}}) \mu}\, ; \label{eq:beta-mu}\\ 
A_{\beta}(x & =\log_{10}({\rm M_{h}})) = A_{0} x^{2} + A_{1} x+ A_{2}\, ;\label{eq:beta-A}\\
B_{\beta}(x & =\log_{10}({\rm M_{h}})) = B_{0} x^{2} + B_{1} x+ B_{2}\, ; \label{eq:beta-B}\\
C_{\beta}(x & =\log_{10}({\rm M_{h}})) = C_{0} x^{2} + C_{1} x+ C_{2}\, . \label{eq:beta-C}
\end{align}
The velocity anisotropy for a fixed halo mass as a function of $\mu$ is simply a power law at the core of the halo with exponential suppression on the outskirt as described by equation~\ref{eq:beta-mu}. The three coefficients $(A_{\beta},B_{\beta},C_{\beta})$ are modelled as quadratic functions of the logarithm of halo mass given by equation~\ref{eq:beta-A},~\ref{eq:beta-B} and ~\ref{eq:beta-C} respectively. The dashed black lines in the middle panel of Figure~\ref{fig:vsat-mu} represents the best-fit model with parameters set to $A_0=0.16,A_1=-4.39,A_2=30.58,B_0=0.20,B_1=-5.45,B_2=37.32,C_0=0.25,B_1=-7.54,B_2=56.35$. This empirical model gives a good description of the velocity anisotropy measured from N-body simulations.

The bottom panel of Figure~\ref{fig:vsat-mu} shows the radial velocity measured from N-body simulation with markers. These velocities are corrected for the Hubble flow and show an amplitude at the level of 5\% of the velocity dispersion in the most massive haloes, indicating that the most massive haloes are very slowly accreting mass. This could imprint a small infall velocity on the satellite galaxies. We propose an empirical model to describe the accretion by a halo as a function of halo mass and mass enclosed fraction parameter ($\mu$):
\begin{equation}
    v_{\rm rad}({\rm M_h},\mu)=v_0 \left[ \log_{10}({\rm M_h})-11.0\right]^{3} (\mu)^{3/2}\, .
    \label{eq:vrad-mu}
\end{equation}
The radial infall velocity goes as the cube of the logarithm of halo mass and as the 3/2 power of the halo mass fraction parameter ($\mu$). $v_0$ is the free parameter that decides the amplitude of radial velocity infall. We obtain $v_0=-2.37$ by fitting the N-body simulations and show the best-fit model with the black dashed line in the bottom panel of Figure~\ref{fig:vsat-mu}.

In our modelling of the GAMA data, we considered applying velocity anisotropy using our empirical fit given in equation~\ref{eq:vrad-mu} to assess the impact of such anisotropy in the measurement and how variations in the anisotropy could be reflected in the small-scale clustering of galaxies. For the purpose of this paper, however, we will ignore such velocity anisotropy effects: the main empirical justification for doing so is that the mass weighted dispersion shows rather little variation with radius, as shown in the top panel of Figure~\ref{fig:vsat-mu}, and therefore the effect of velocity anisotropy is largely included when we allow the satellite velocity dispersion to vary.

\subsubsection{Satellite galaxies: density distribution vs velocity dispersion}
\label{sec:sat-rho-vrms}

\begin{figure}
\centering
\includegraphics[width=0.48\textwidth]{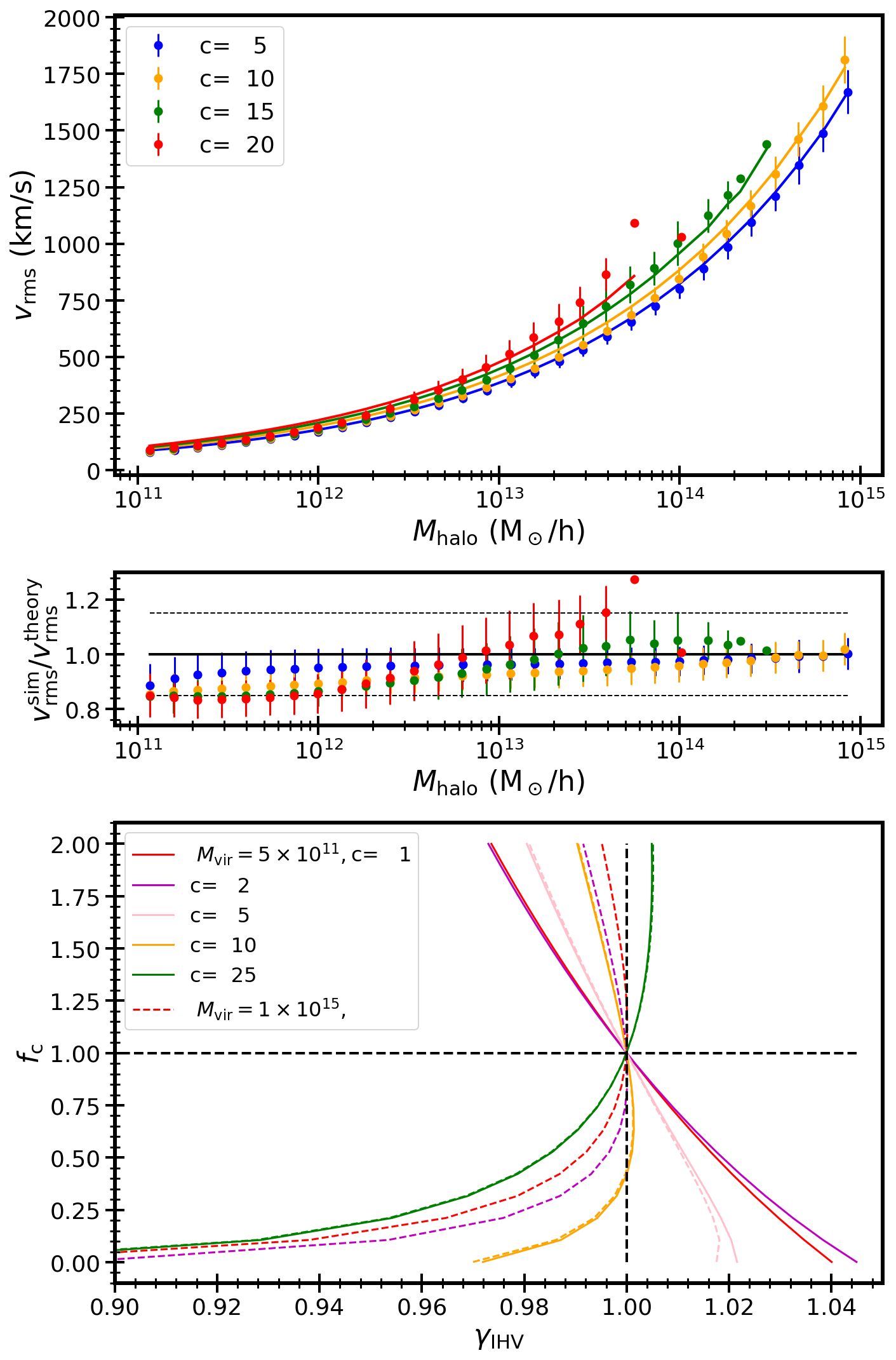}
    \caption{The internal velocity dispersion of dark matter haloes and its relation to the concentration of the satellite galaxies. The top panel shows the velocity dispersion of dark matter haloes with halo mass for different bins of concentration. The circles denotes measurements from numerical halo catalogues whereas the lines show analytical predictions assuming spherically symmetric NFW haloes. The agreement between lines and points indicates that the velocity dispersion is not greatly impacted by deviation from spherical symmetry. The middle panel shows the ratio between the velocity dispersion measured from N-body simulations and the theoretical prediction. The bottom panel shows the constraint relation between $f_c$ and $\gamma_{\rm IHV}$ for two different halo masses with solid and dashed lines and for several values of halo concentration, shown in different colours.}
    \label{fig:Halo_fc_gIHV}
\end{figure}

One can argue that the the efficiency of galaxy formation within a halo can easily be a function of distance. Such variation in efficiency will result in the satellite galaxies distribution being different from the distribution of dark matter itself within a halo. It is interesting to ask how such a phenomenon would affect the velocity dispersion of satellites and hence the  small-scale clustering in redshift space. In order to do so we first consider spherically symmetric NFW haloes and theoretically evaluate the mass-weighted velocity dispersion as a function of halo mass and concentration. We compare the theoretical estimate of velocity dispersion with the one measured from numerical halo catalogues. This is shown in the top panel of Figure \ref{fig:Halo_fc_gIHV}. In the middle panel we show the ratio of velocity dispersion measured in N-body simulations with the theoretical estimates as a function of halo mass and concentration. The theoretical prediction seems to work reasonably well except in highly concentrated or low mass haloes, where we see an error of the order of $15\%$. This estimate of velocity dispersion is good enough for our study as we will be using it only to scale the velocity dispersion estimated from N-body simulations. 

We allow for the mass distribution of the satellite galaxies being possibly different from the dark matter by introducing a single parameter $f_c$, which is the ratio of the scale radii of the two distributions (see equation~\ref{eqn:rho0}). We then compute the ratio $\gamma_{\rm IHV}$ between the velocity dispersion of satellite galaxies with a given distribution to the velocity dispersion of satellite galaxies that trace the mass. Note that we implicitly assume here that the masses of the galaxies themselves are negligible, so that the velocity dispersion profile is given only by the dark matter distribution in the halo. The result of this calculation is a constraint relation between $f_c$ and $\gamma_{\rm IHV}$ as shown in the bottom panel of Figure \ref{fig:Halo_fc_gIHV} for a given halo mass and concentration. The lines with different colours show the effect of changing concentration whereas solid and dashed lines represent the two different halo masses of $5\times 10^{11}\msolaroh$ and $1\times 10^{15} \msolaroh$. 

We note that the constraint relation ($f_c \leftrightarrow \gamma_{\rm IHV}$) strongly depends on the concentration. For haloes with lower dark matter concentration $f_c$ and $\gamma_{\rm IHV}$ are negatively correlated whereas for more concentrated haloes they are positively correlated. We also note that the constraint relation is independent of the the mass of the halo for haloes with concentration above 5. But for haloes with lower concentration the constraint relation show a mass dependence, and the solid and dashed lines are different in the bottom panel of Figure \ref{fig:Halo_fc_gIHV}. 

It is also interesting to note that the velocity dispersion of satellite galaxies is hard to change even when we modify the galaxy concentration by large amount. The change in velocity dispersion stays within  $5\%$ except when we reduce the concentration of the satellite galaxies by a factor of 10 (i.e. $f_c<0.1$) and the halo has concentration $c>10$; in this case it shows more than $10\%$ reduction in the velocity dispersion of the satellite galaxies. This is also good news for dynamical masses, implying systems with hydrostatic equilibrium will have quite stable velocity dispersion even if we use a relatively biased distribution of satellite galaxies. We therefore use this constraint relation in our analysis where we set $\gamma_{\rm IHV}$ for a given haloes based on its virial mass and concentration for a fixed value $f_c$. Note that when using the constraint relation $f_c$ is still a single parameter independent of any halo properties, but $\gamma_{\rm IHV}$ now depends on halo properties, namely mass and concentration.

\subsubsection{Satellite galaxy radius}
One more uncertainty regarding satellite galaxy distribution around haloes is the question of the halo boundary as regards satellite galaxies. Typically, one assumes that the satellite galaxies are truncated at the halo boundary defined as the virial radius. But the NFW density profile formula lacks a sharp truncation and there is no clear reason for satellite galaxies to be precisely limited by the virial radius of the mass. Therefore, we consider the question of how far one should extend the satellite galaxies, and whether data can tell us anything about such freedom. This motivates us to introduce an additional parameter $f_{\rm vir}$, which is the maximum distance out to which satellite galaxies should be populated around haloes in units of the virial radius: i.e. we populate satellite galaxies up to a distance of $f_{\rm vir}r_{\rm vir}$ around the halo centre. This also implies that the velocity dispersion of satellite galaxies will depend on $f_{\rm vir}$. To account for this change in velocity dispersion we analytically estimate the ratio of mass-weighted velocity dispersion at a distance of $f_{\rm vir} r_{\rm vir}$ and $r_{vir}$. This analytical estimate of the ratio of velocity dispersion is then used to scale the numerically estimated velocity dispersion of each halo within the virial radius. The resulting scaled velocity dispersion is then assigned to the satellite galaxies.

\subsection{Impact of cosmological parameters}

One issue for RSD analysis is that the background cosmological model is not known exactly. In practice, a fiducial cosmology is assumed in order to convert observables of angle on the sky and redshift into comoving spatial separations:
\begin{equation}
\Delta r_\perp = (1+z)D_A(z)\, \Delta\theta; \quad
\Delta r_\parallel = c\, \Delta z/H(z) \, ,
\end{equation}
where $D_A$ is the usual proper angular-diameter distance. If the fiducial cosmology is not correct, both $D_A$ and $H$ will alter, but not by the same factor, leading to anisotropic clustering. Since the fiducial cosmology is known by definition, a measurement of this apparent anisotropy will yield an estimate of the parameter
\begin{equation}
F(z)\equiv (1+z)D_A(z)H(z)/c\, .
\end{equation}
This signature was first proposed by 
as a means of constraining the expansion history via $F(z)$ 
\citep{1979Natur.281..358A}, but the measurement of Alcock--Paczynski distortion is nearly degenerate with RSD anisotropy \citep{1996MNRAS.282..877B}. We allow the allow the freedom of Alcock--Paczynski parameters in our model for some of the analysis to asses its degenracy with RSD anisotropy and impact on parameter constrinats.

\section{Measurements}
\label{sec:measurement}

\begin{figure*}
    \centering
    \includegraphics[width=0.8\textwidth]{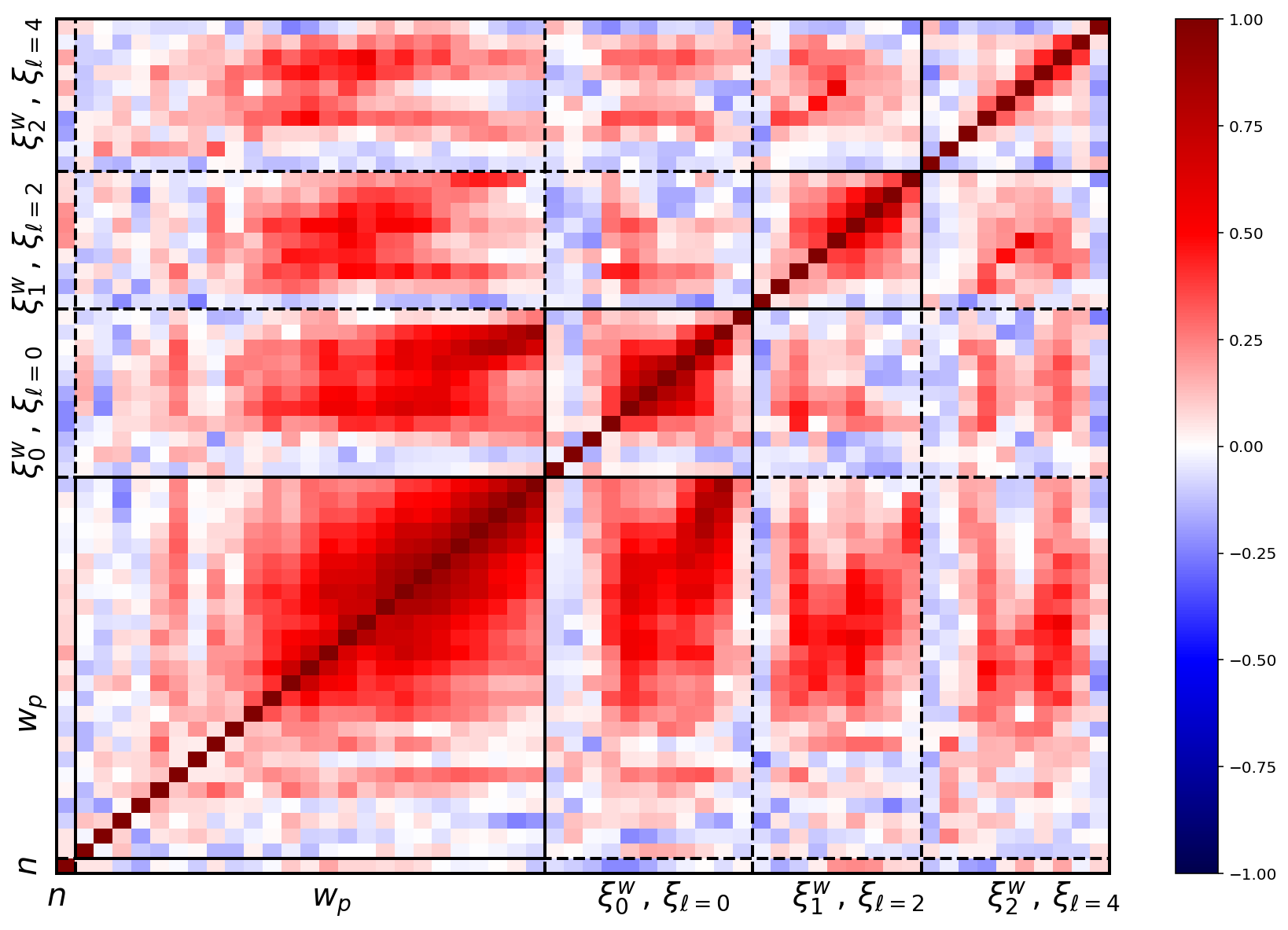}
    \includegraphics[width=0.8\textwidth]{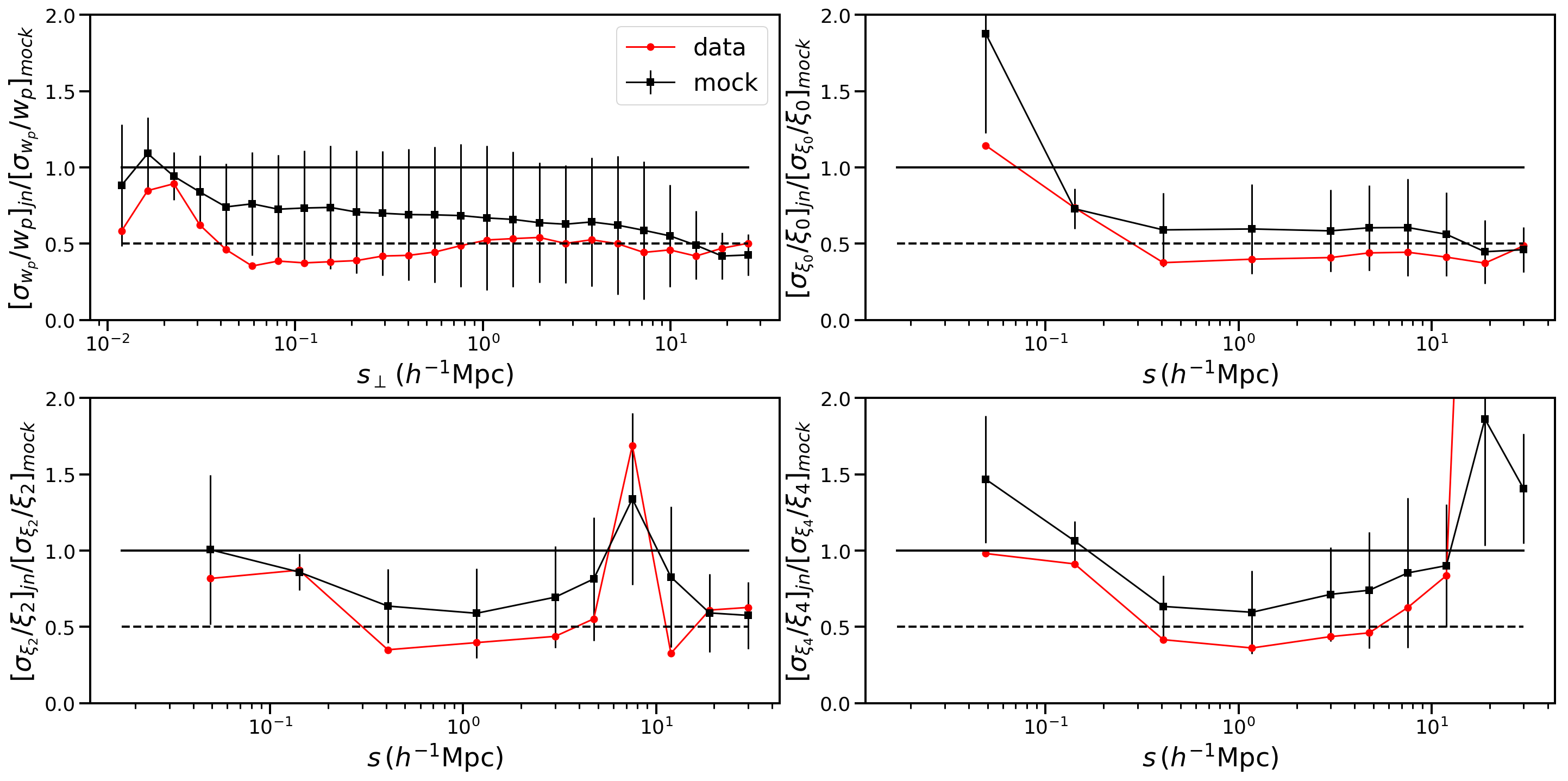}
    \caption{Correlation matrix and relative diagonal error on our observable measured using combination of jackknife and mocks. The top plot shows the correlation matrix between $w_p$ , $\hat{\xi}_0$, $\hat{\xi}_2$ and $\hat{\xi}_4$ estimated using 108 jackknife regions for the $M_r<-21$ subsample of GAMA. The bottom four panels shows the relative error estimated using jackknife for both mocks and data with respect to error estimated from the variance of mock. The jackknife errors are typically a factor 2 smaller compared to the mock error at all scales.}
    \label{fig:corrmat}
\end{figure*}

We start with four measured quantities for each galaxy in our sample: right ascension (RA), declination (dec), redshift ($z$) and extinction corrected $r$-band Petrosian magnitude $r_{\rm petro}$. We first estimate the $r$-band absolute magnitude ($M_r$) using the following equation:
\begin{equation}
\eqalign{
    M_r -5 \log_{10}{h}= &r_{\rm petro} -5 \log_{10}{\f{D_L(z)}{\mpcoh}} -25 \cr 
    &-k_{\rm col}(z) -E(z,M)\, ,
    }
    \label{eqn:magnitude}
\end{equation}
where $D_L(z)$ is the luminosity distance, $k_{\rm col}(z)$ is the colour dependent k-correction derived in \cite{2014MNRAS.445.2125M} and $E(z,M)$ is the magnitude dependent evolution correction described in section \ref{sec:data}.
We then select three absolute magnitude limited samples with $M_r<-21$, $M_r<-20$ and $M_r<-19$, giving us volume limited samples out to maximum redshifts of 0.36, 0.26 and 0.17 respectively. The maximum redshift for subsamples were obtained after applying the k+e correction to the survey flux limit. The volume occupied by the three samples are 0.0175, 0.0071 and 0.0021 in units of $(\mathrm{h}^{-1}\mathrm{Gpc})^3$ for $M_r<-21$, $M_r<-20$ and $M_r<-19$ respectively. We also apply the same procedure to the random catalogue provided with the GAMA data and generate matching random samples \cite{2015MNRAS.454.2120F}. In order to estimate errors and the covariance matrix we split the survey into 108 jackknife regions by dividing each of the 12$\times$5 deg$^2$ GAMA fields into 9$\times$4 regions. This allows 108 realisations for any measurement by removing one jackknife region each time. We then measure the galaxy number density and clustering with their covariances as described below.

\subsection{Galaxy Number density}
In order to estimate the number density we first need to estimate the area of the survey after removing all masked regions. We used a mask of the survey in the form of a Healpix map with $N_{\rm side}=2048$. We then use $10^8$ randoms uniformly distributed over the full sky to count the number of pixels in the survey ($n_{\rm survey}$). Therefore the number density ($n_{\rm jn}$) for each jackknife realisation is given by
\begin{eqnarray}
n_{\rm jn} &=& \f{n_{\rm galaxy}^{\rm jn}}{\f{4}{3} \pi \left(\chi^3(z=z_{\rm max})-\chi^3(z=0.05)\right) A_{\rm jn}}  \\
A_{\rm jn} &=& \f{360^2}{\pi} \f{n_{\rm survey}^{\rm jn}}{10^8}\, ,
    \label{eqn:density}
\end{eqnarray}
where $n_{\rm galaxy}^{\rm jn}$ and $n_{\rm survey}^{\rm jn}$  are respectively the number of galaxies and the number of randoms included in the survey for given jackknife realisation ${\rm jn}=1-108$. $\chi(z)$ is the comoving distance at redshift $z$ in $\mpcoh$ units.

\subsection{Correlation function}
We first estimate the position of each galaxy in 3-dimensional space by converting redshift to line-of-sight distance using our fiducial cosmology.
We then measure the galaxy auto-correlation function using the minimum variance Landay-Szalay estimator \citep{LandySzalay93} given by:
\begin{equation}
    \xi(\vec{r})=\f{DD(\vec{r})-2DR(\vec{r})+RR(\vec{r})}{RR(\vec{r})}\, ,
    \label{eqn:landay-szalay}
\end{equation}
where DD, DR and RR are the numbers of galaxy-galaxy, galaxy-random and random-random pairs as a function of the difference vector in 3-dimensional space. We note that redshift space distortions make the line-of-sight a special direction, and we therefore project the 3-dimensional space onto a 2-dimensional space that decomposes pair separation vectors along line-of-sight ($s_{\parallel}$) and perpendicular to the line-of-sight ($s_{\perp}$). This gives us a 2-dimensional correlation function $\xi(s_\parallel,s_\perp)$. 

We first measure the projected correlation ($w_p$) by integrating the 2-dimensional correlation function along the line-of-sight between $s_\parallel=-40\mpcoh$ to $s_\parallel=+40\mpcoh$ (following appendix B of \citet{2018MNRAS.474.3435L}) and using 25 logarithmically spaced bins in $s_\perp$ between $0.01\mpcoh$ and $30\mpcoh$. The projected correlation function helps us constrain the HOD parameters that govern the galaxy-halo connection. In order to measure the redshift space distortion and impact of galaxy velocity we need to quantify the shape of the anisotropic galaxy clustering in this 2-dimensional space. We measure the anisotropy of galaxy clustering using two different estimators at very small ($s<2\mpcoh$) and small scales ($2\mpcoh<s<30\mpcoh$) to overcome the limitation of our measurements. We first measure 3 `wedge' statistics covering below $2\mpcoh$ as follows:
\begin{equation}
\xi(s)_{w}^{\rm wedge}=\int_{\mu=w/3}^{\mu=(w+1)/3} \xi(s,\mu)\, d\mu\, ,
    \label{eqn:wedge}
\end{equation}
where $s$ is the pair separation in $\mpcoh$ and $\mu=\cos(\theta)$ where $\theta$ is the angle of pair separation from line-of-sight. The $w$ takes values 0,1,2 giving us three wedges $\xi(s)_{0}^{\rm wedge},\xi(s)_{1}^{\rm wedge}$ and $\xi(s)_{2}^{\rm wedge}$ respectively. We create 5 logarithmic bins in wedges between $s=0.01\mpcoh$ and $s=2\mpcoh$.
We note that $0.01\mpcoh$ corresponds to 2 arc second at z = 0.36 (the upper redshift limit of our sample), hence caution is needed to push measurements below this scale, as the target sample may encounter blending-related issues in dense regions. We measure multipole moments of the galaxy clustering for scale above $2\mpcoh$. The Legendre decomposition is given as follows:
\begin{equation}
\xi(s)_{\ell}=\int_{\mu=0}^{\mu=1} \xi(s,\mu) \mathcal{P_{\ell}}(\mu)\, d\mu\, ,
    \label{eqn:multipole}
\end{equation}
where $\mathcal{P_{\ell}}(\mu)$ is the Legendre polynomial of order $\ell$ with $\ell=0,2,4$ corresponding to monopole, quadrupole and hexadecapole respectively. We create 6 logarithmically spaced bins of multipoles in $s$ between $3\mpcoh$ and $30\mpcoh$. we distinguish between wedges at very small scales and multipoles at small scales in order to perform these measurements in an efficient way without introducing convergence issues in any integral. In principle we should be able to measure multipoles at all scales which is our default choice. But at very small scales in order to measure multipole moments we need a large number of bins in $\mu$ for the integral in equation \ref{eqn:multipole} to converge. But this will mean several of our bins might have zero counts at very small scales which can lead to issues in measurement. Therefore we avoid this by making wide $\mu$ bins and working with wedges at these smallest scales. We denote the combination of wedges at small scales and multipoles at large scales as follows:
%
\begin{align}
\hat{\xi}_\ell(s) =[ & \xi_0^{\rm wedges}(s<2\mpcoh),\xi_{\ell=0}(s>2\mpcoh) \\
 & \xi_1^{\rm wedges}(s<2\mpcoh), \xi_{\ell=2}(s>2\mpcoh) \nonumber \\
 & \xi_2^{\rm wedges}(s<2\mpcoh), \xi_{\ell=4}(s>2\mpcoh) ] \, . \nonumber 
\label{eqn:xihat}
\end{align}

\subsection{Covariance matrix}

We estimate the full covariance matrix using jackknife and mocks together. We first divide each of the three GAMA fields into 9$\times$4 jackknife regions, which gives us altogether 108 jackknife regions for our data. We then measure all of our observables by removing one jackknife region at a time leading to a total of 108 jackknife realisations. We compute the covariance matrix from these 108 realisation. The top plot in Figure \ref{fig:corrmat} shows the correlation matrix for auto and cross terms of number density, $w_p$ , $\hat{\xi}_0$, $\hat{\xi}_2$ and $\hat{\xi}_4$ estimated from this jackknife procedure. Such an estimate of the covariance matrix might underestimate the diagonal error as it will not capture the cosmic variance terms entirely. Therefore we measure the diagonal error on all of our observables using 24 mocks from \citet{2013MNRAS.429..556M}. We then compare the diagonal error estimated from the mocks with the diagonal error estimated from the jackknife in the bottom panel of Figure \ref{fig:corrmat}. The black points shows the ratio of jackknife error and mock error where the error bars are estimated by estimating jackknife errors for each of the mocks and looking at the variance in the ratio of jackknife error to the mock error over all the mocks. The red points represents the ratio of jackknife error on data to the mock error. We find that the jackknife error on the data is smaller at most by a factor of 2 compared to the mock errors. Our finding is consistent with results reported in Figure 6 of \cite{2015MNRAS.454.2120F}. Therefore we multiply our jackknife covariance matrix by a factor of 4 which should at best over-estimate our error bars. We also note that the theoretical predictions will also have error due to limited volume of simulation used. But the volume of survey is much smaller than simulation used except for $M_r<-21$ sample. Given our errors are conservative we do not add any additional contribution to the covariance matrix coming from theoretical uncertainty. A better estimate of the covariance matrix is desirable but outside the scope of this work. In fact this work should help us improve the estimate of the covariance matrix for future analysis by providing an improved model at all scales. By applying our final model to several realisations of the N-body simulation we should be able to construct a better covariance matrix in the deeply non-linear scale used in this analysis.

\begin{table}
    \centering
    \begin{tabular}{|p{1.2cm}|p{4.5cm}|p{1.5cm}|} \hline\hline
    Parameters &  Description & prior \\ [0.5ex] \hline\hline 
    $M_{\rm cut}$ & Halo mass at which probability of having central galaxy is 0.5.  & $10^{11}$--$10^{15}$ \\ 
    $\sigma_{M}$ & scatter in the halo mass to model the given absolute magnitude limited sample. This should be related to scatter in halo mass and absolute magnitude of galaxies. & 0--8\\
    $\kappa$ & This determines the mass at which haloes have no satellite galaxies in units of $M_{\rm cut}$ & 0--3\\
    $M_{1}$ & This determines the scaling of number of satellite galaxies with halo mass. &  $10^{11}$--$10^{15}$\\
    $\alpha$ & The power law index of number of satellite as the function of halo mass. & 0--3 \\ \hdashline
    $f_c$ & The distribution of galaxies might follow a different concentration than dark matter itself. This parameter scales the concentration of the dark matter halo to determined the concentration of the satellite galaxies by scaling the scale radius $R_s$ of the halo. & $10^{-3}$--5 \\
    $\gamma_{\rm HV}$ & This parameters scales the inter-halo velocity to allow an additional degree of freedom as the growth rate of structure. & 0--3\\
    $\gamma_{\rm IHV}$ & This scales the velocity dispersion of the dark matter halo in order to allow the satellite galaxy velocity distribution to be different from dark matter. & 0--3\\ 
    $f_{\rm vir}$ & This scales the maximum distance up to which satellite galaxies are distributed in unit of virial radius of the halo. A corresponding velocity dispersion is also estimated based on the according to the distance. & 0.1--5\\ \hdashline
    $\alpha_{\parallel}$ & Alcock-Paczynski effect by scaling the line-of-sight distances. & 0.5--1.5 \\
    $\alpha_\perp$ & Alcock-Paczynski effect by scaling the distances perpendicular to the line-of-sight. & 0.5-1.5\\ \hline\bottomrule
    \end{tabular}
    \caption{The list of model parameters with their short descriptions and prior range used in our analysis.}
    \label{tab:parlist}
\end{table}

\subsection{Analysis methods}
\label{sec:analysis}

We start with the halo catalogue produced using the ROCKSTAR halo finder on the Bolshoi N-body simulation. In the halo catalogue we have measurements of concentration, virial radius, scaled radius, halo mass, core velocity and velocity dispersion as described in section \ref{sec:model}.  For each point in the parameter space we first assign the number of central and satellite galaxies using equation \ref{eqn:HOD}. We then place the central galaxy at the centre of the halo and assign a velocity equal to the halo's core velocity, but scaled by a free parameter $\gamma_{\rm HV}$. For satellite galaxies we assign positions that follow an NFW profile for the given halo based on its concentration, mass and radius parameters where in all cases the galaxy concentrations is scaled with respect to the intrinsic halo concentration by a global free parameter $f_c$. We then assign satellite velocities sampled from a Gaussian distribution with the same mean as the central galaxy and a dispersion given by the halo velocity dispersion multiplied by a free parameter $\gamma_{\rm IHV}$. We then move each galaxy to its redshift-space position assuming that the line-of-sight direction lies along the $z$-axis of the periodic box, adopting the plane parallel approximation. We now treat this dataset as an observed catalogue and measure all of our observables, namely number density, $w_p$, $\hat{\xi}_0$, $\hat{\xi}_2$ and $\hat{\xi}_4$. These results are then used as our model prediction. The parameter space is then sampled to estimate the posterior distribution of our parameters given the data and covariance matrix.  
In practice, we perform this sampling via an MCMC analysis using {\tt emcee}.

The full list of parameters with their descriptions and priors is given in Table \ref{tab:parlist}.

\section{Results}
\label{sec:result}

\begin{figure*}
    \centering
    \includegraphics[width=1.0\textwidth]{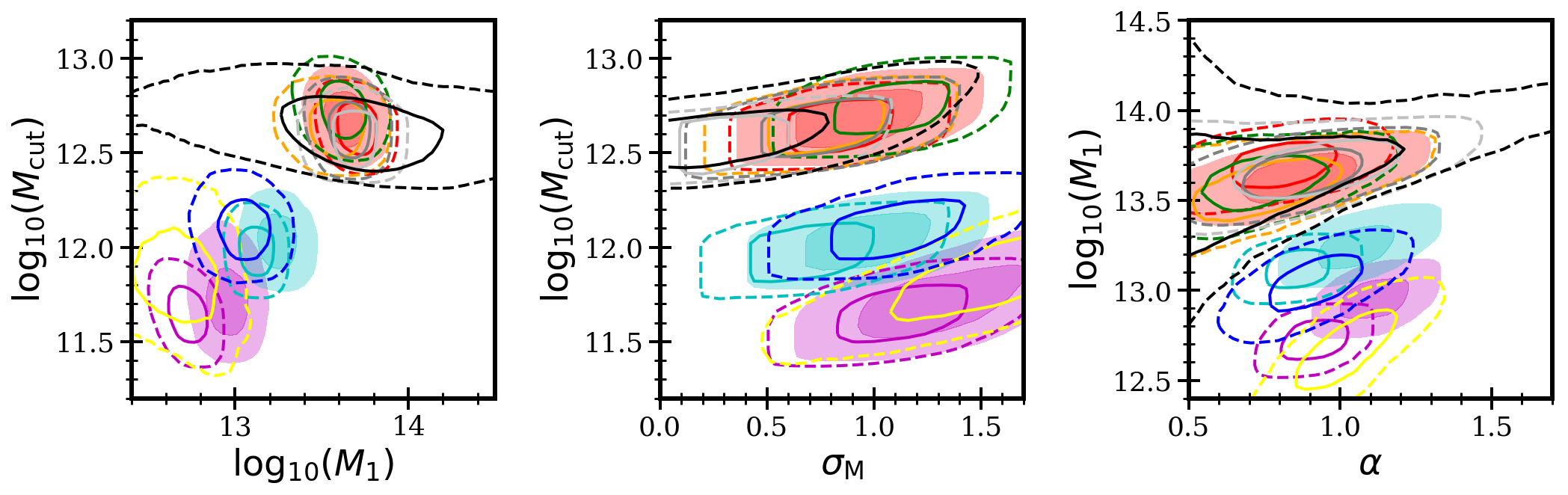}
    \includegraphics[width=1.0\textwidth]{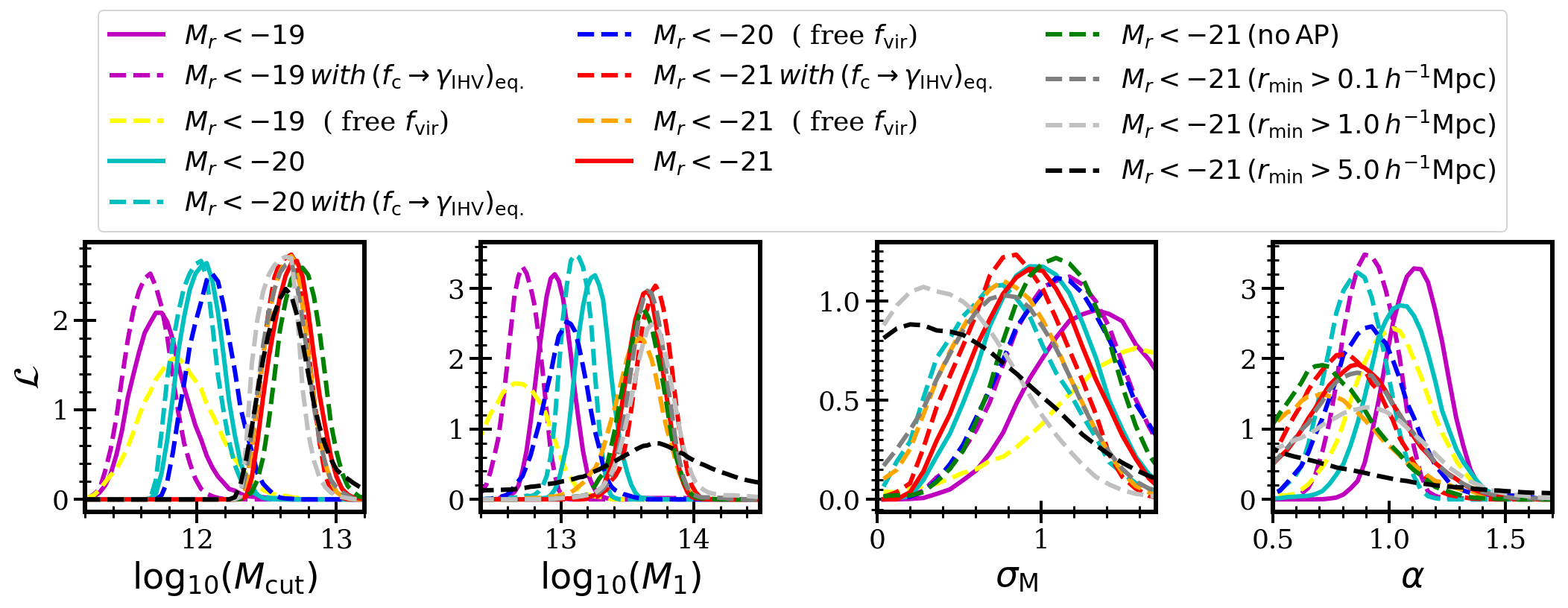}
    \caption{The two dimensional and one dimensional constraints on the base HOD parameters for three r band absolute magnitude limited ($M_r<[-21,-20,-19]$) subsamples in GAMA. The solid and dashed contours of given colour represents the 1$\sigma$ and 2$\sigma$ confidence limits respectively. The yellow unfilled, magenta filled and magenta unfilled contours represents the constraints for $M_r<-19$ subsample with all three satellite parameters ($f_c,\gamma_{\rm IHV},f_{\rm vir}$) free, with fixed $f_{\rm vir}$ and with fixed $f_{\rm vir}, \gamma_{\rm IHV}$ using constraint ($f_c \rightarrow \gamma_{\rm IHV}$) relation respectively. Similarly, the blue and cyan contours represent the three cases for the $M_r<-20$ subsample. Finally, orange and red contours are for $M_r<-21$. The green contours are when Alcock--Paczyinski parameters are also fixed for the $M_r<-21$ subsample. The grey, silver and black contours are for the $M_r<-21$ subsample with different minimum scale of $0.1\mpcoh$, $1.0\mpcoh$ and $5.0\mpcoh$ respectively.}
    \label{fig:HODbase}
\end{figure*}

\begin{figure*}
    \centering
    \includegraphics[width=1.0\textwidth]{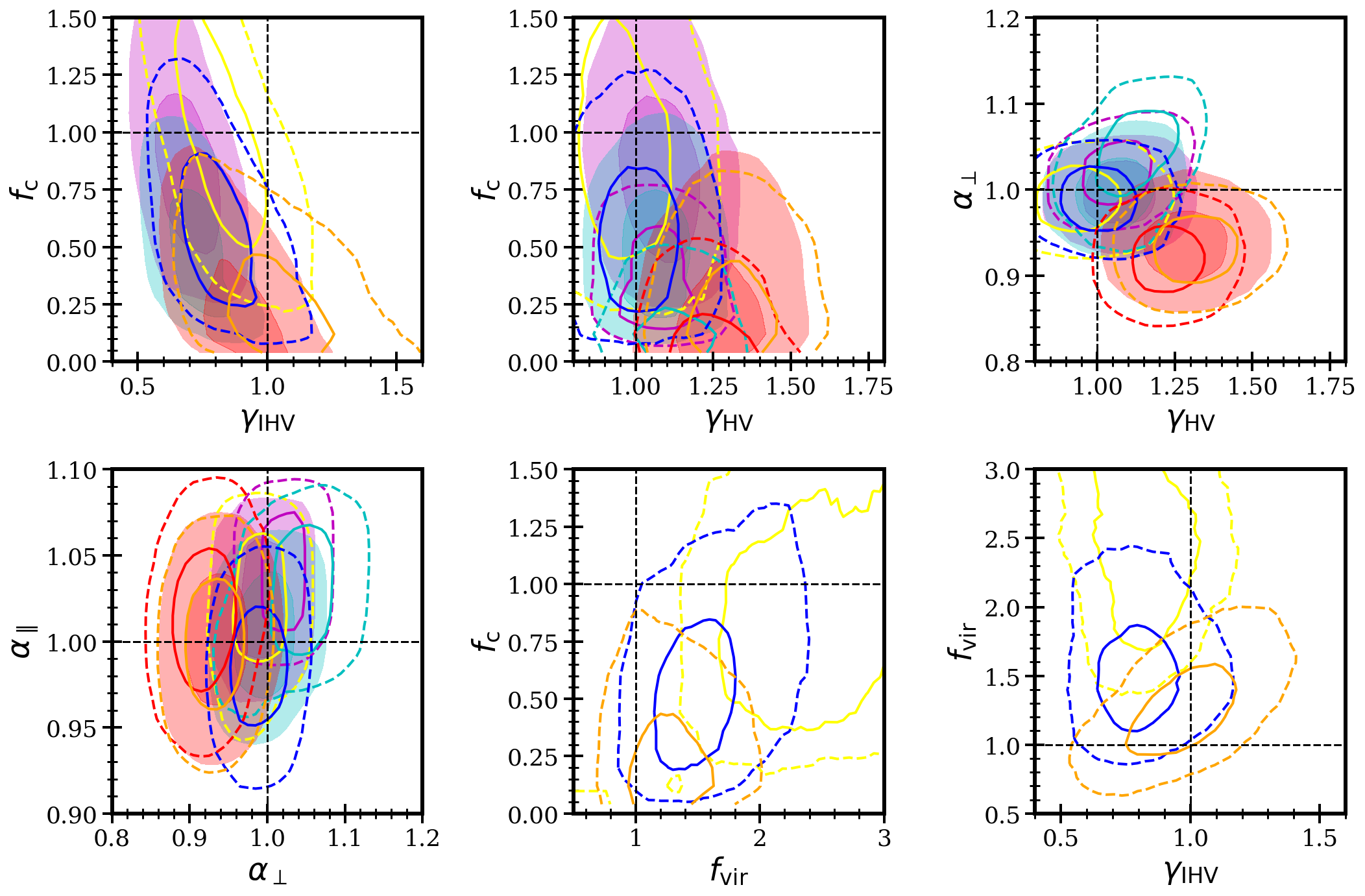}
    \includegraphics[width=1.0\textwidth]{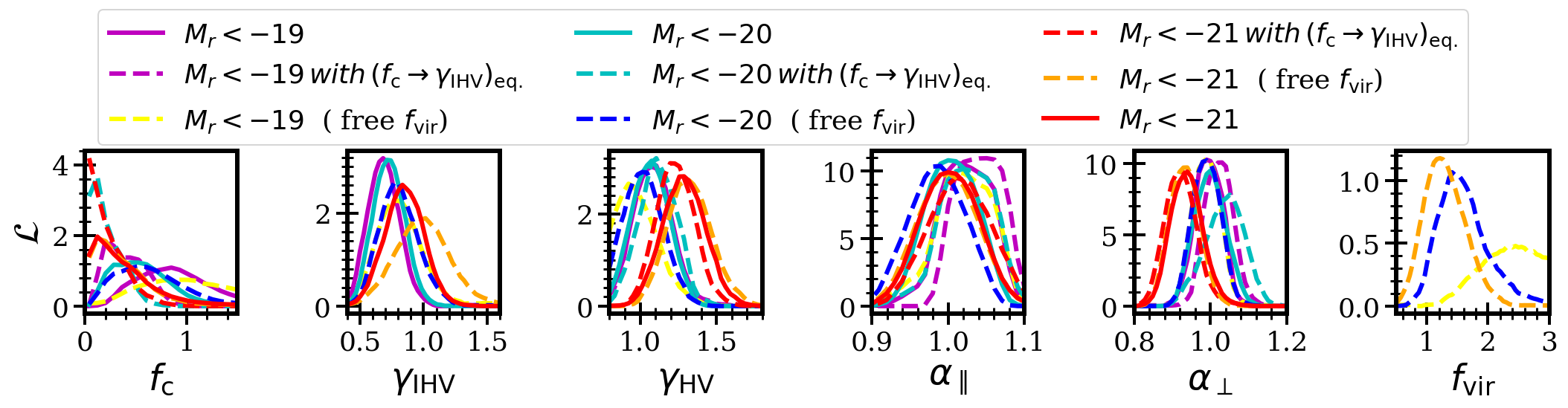}
    \caption{The same as Figure~\ref{fig:HODbase}, but going beyond the base HOD parameters. The top left panel shows that if we allow satellites to have different concentration and velocity dispersion compared to dark matter haloes, then the brighter ($M_r<-21$) satellites show the same velocity dispersion but much lower concentrations, whereas including the fainter satellites ($M_r<-19$) increases the concentration of the satellite distribution to be consistent with dark matter but requires a smaller velocity dispersion. If we allow the maximum radii of satellites to be free then the velocity dispersion of fainter satellite becomes more consistent with the dark matter, although the same trend for concentration is still seen.}
    \label{fig:HODextended}
\end{figure*}

We present one of the first constraints on the HOD parameters for absolute magnitude limited samples using GAMA. 
Figure \ref{fig:HODbase} shows the two- and one- dimensional constraints on four of the five base HOD parameters that connect halo populations to the galaxy population, using three absolute magnitude limited subsamples $M_r<-21,-20,-19$ in GAMA. The only parameter left unconstrained by the data, not shown in the plot, is $\kappa$. Figure~\ref{fig:HODextended} similarly shows the two and one dimensional constraints on the extended set of HOD parameters. In the two dimensional posterior plots (in Figure~\ref{fig:HODbase} and~\ref{fig:HODextended}),  the solid and dashed contours of a given colour represent the 1$\sigma$ and 2$\sigma$ confidence limits respectively. Note that there are three parameters ($f_{\rm vir}, \gamma_{\rm IHV}$ and $f_c$) that set the satellite distribution and dynamics. The yellow unfilled, magenta filled and magenta unfilled contours represent the constraints with all three satellite parameters free, with fixed $f_{\rm vir}$ and with fixed $f_{\rm vir}, \gamma_{\rm IHV}$ using the constraint relation for the $M_r<-19$ subsample. Similarly, the blue unfilled, cyan filled and cyan unfilled contours represents the three cases for the $M_r<-20$ subsample. Finally, orange unfilled, red filled and red unfilled contours are for $M_r<-21$. The green unfilled contours are when Alcock--Paczyinski parameters are also fixed for the $M_r<-21$ subsample. The grey, silver and black contours (only in Figure~\ref{fig:HODbase}) are for the $M_r<-21$ subsample with different minimum scales of $0.1\mpcoh$, $1.0\mpcoh$ and $5.0\mpcoh$ respectively. We note that the base HOD parameters are insensitive to the details of the satellite degrees of freedom. We also show for the $M_r<-21$ sample that allowing extra degree of freedom with the Alcock--Paczynski parameter does not affect the HOD parameters constraint, and hence we are also relatively immune to the details of the fiducial cosmology assumed in the analysis. Another interesting question to ask is what scales are relevant for the purpose of constraining the base HOD model. We show that scales below $1\mpcoh$ have negligible information about the base HOD parameters. We also show that including much smaller scales with a  variety of assumption about small scale satellite population does not bias the base HOD parameters. For all subsamples we use scales from $0.01\mpcoh$ to $30\mpcoh$, covering more than three orders of magnitudes, from deep inside haloes to the cosmological linear scales. Table~\ref{tab:HODMr21_wscale} shows the constraints on various parameters for the $M_r<-21$ sample with different choices of scales. We show the rest of the results on parameter constraints for different subsamples and with various model assumptions in table~\ref{tab:HOD_wMr}. 

\begin{figure*}
    \centering
    \includegraphics[width=0.98\textwidth]{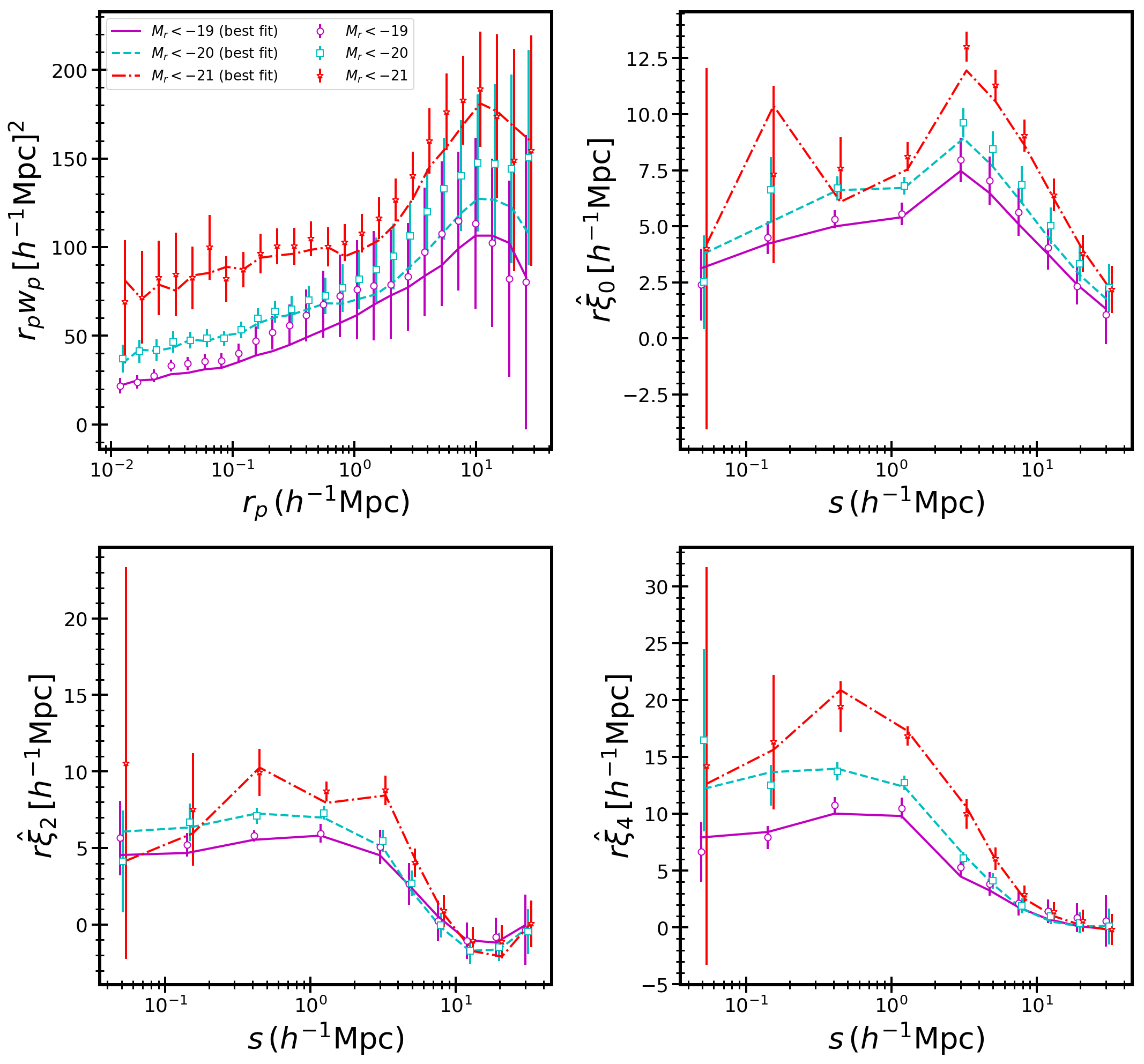}
    \caption{Clustering measurements for $M_r<-19$, $M_r<-20$ and  $M_r<-21$ subsamples in GAMA, together with best fit model. The measurements are shown with points and jackknife errors. We note that the first four bins plotted in the multipole plots are actually wedges. The best fit models are shown with lines as indicated in the legend.
    The $w_p$, $\hat{\xi}_0$, $\hat{\xi}_2$ and $\hat{\xi}_4$  are shown in the top left, top right, bottom left and bottom right panels respectively. We only show the best model lines corresponding to our fiducial model.}
    \label{fig:wpxis024_Mr21}
\end{figure*}

We show the measured clustering statistics $w_p$, $\hat{\xi}_0$, $\hat{\xi}_2$ and $\hat{\xi}_4$ in Figure \ref{fig:wpxis024_Mr21} including the best fit models with different minimum fitting scales. Our model shows consistent prediction of the data to the smallest scales.

\begin{table*}
\strut\bigskip
\setlength{\belowrulesep}{-0.3em}
	\centering
	\begin{tabular}{|c| c| c| c| c| c|} 
	\toprule
	 \medstrut{\bf Parameters} & $\mathbf{s_{\rm \bf min}=0.01}$ & $\mathbf{s_{\rm \bf min}=0.01}$ (no AP) & $\mathbf{s_{\rm \bf min}=0.10}$ & $\mathbf{s_{\rm \bf min}=1.0}$ & $\mathbf{s_{\rm \bf min}=5.0}$ \\ [0.8ex] 
	 \multicolumn{6}{|l|}{\bf Basic HOD model} \\ \hdashline
	$\log_{10}(M_{\rm cut})$ & $12.68^{+0.11}_{-0.08}$  & $12.74^{+0.1}_{-0.09}$  & $12.63^{+0.1}_{-0.08}$  & $12.57^{+0.09}_{-0.05}$  & $12.64^{+0.16}_{-0.1}$  \\ [0.9ex]
	$\sigma_{\rm M}$ & $0.93^{+0.27}_{-0.26}$  & $1.09^{+0.21}_{-0.23}$  & $0.77^{+0.29}_{-0.34}$  & $0.47^{+0.36}_{-0.3}$  & $0.57^{+0.53}_{-0.38}$  \\ [0.9ex]
	$\log_{10}(M_1)$ & $13.63^{+0.11}_{-0.1}$  & $13.62^{+0.12}_{-0.13}$  & $13.66^{+0.1}_{-0.11}$  & $13.69^{+0.13}_{-0.14}$  & $13.71^{+0.61}_{-0.64}$  \\ [0.9ex]
	$\alpha$ & $0.86^{+0.2}_{-0.21}$  & $0.72^{+0.21}_{-0.19}$  & $0.86^{+0.21}_{-0.21}$  & $0.84^{+0.27}_{-0.36}$  & $0.39^{+0.65}_{-0.28}$  \\ [0.9ex]
	$\kappa$ & $2.17^{+0.54}_{-0.87}$  & $2.14^{+0.6}_{-0.84}$  & $2.16^{+0.6}_{-0.85}$  & $1.84^{+0.81}_{-1.05}$  & $1.65^{+0.95}_{-1.09}$  \\ [0.9ex]
	 \multicolumn{6}{|l|}{\bf Dynamics and Satellite} \\ \hdashline
	$\gamma_{\rm HV}$ & $1.29^{+0.11}_{-0.12}$  & $1.29^{+0.14}_{-0.12}$  & $1.33^{+0.13}_{-0.12}$  & $1.3^{+0.12}_{-0.13}$  & $0.98^{+0.78}_{-0.73}$  \\ [0.9ex]
	$f_{\rm c}$ & $0.26^{+0.32}_{-0.16}$  & $0.26^{+0.26}_{-0.16}$  & $0.19^{+0.34}_{-1.13}$  & $2.11^{+1.94}_{-1.5}$  & $2.57^{+1.65}_{-1.7}$  \\ [0.9ex]
	$\gamma_{\rm IHV}$ & $0.85^{+0.12}_{-0.14}$  & $0.84^{+0.12}_{-0.14}$  & $0.86^{+0.13}_{-0.13}$  & $1.08^{+0.28}_{-0.26}$  & $1.06^{+0.74}_{-0.48}$  \\ [0.9ex]
	 \multicolumn{6}{|l|}{\bf Alcock--Paczynski parameters} \\ \hdashline
	$\alpha_{\parallel}$ & $1^{+0.03}_{-0.03}$  &  fixed   & $0.99^{+0.03}_{-0.03}$  & $1^{+0.03}_{-0.03}$  & $1.01^{+0.05}_{-0.05}$  \\ [0.9ex]
	$\alpha_{\perp}$ & $0.94^{+0.04}_{-0.03}$  &  fixed   & $0.93^{+0.03}_{-0.03}$  & $0.95^{+0.05}_{-0.03}$  & $1.01^{+0.12}_{-0.09}$  \\ [0.9ex]
	 \multicolumn{6}{|l|}{\bf Inferred Parameters} \\ \hdashline
	$f_{\rm sat}$ & $0.15 \pm 0.02$  & $0.14 \pm 0.01$  & $0.15 \pm 0.02$  & $0.17 \pm 0.03$  & $0.22 \pm 0.11$  \\ [0.9ex]
	$10^{3} \bar{n} \, [\mpcoh]^{-3}$ & $1.61$  & $1.61$  & $1.61$  & $1.61$  & $1.60$  \\ [0.9ex]
	$f\sigma_8(z_{\rm mean})$ & $0.60^{+0.05}_{-0.06}$  & $0.60^{+0.07}_{-0.06}$  & $0.62^{+0.06}_{-0.06}$  & $0.60^{+0.06}_{-0.06}$  & $0.46^{+0.36}_{-0.34}$  \\ [0.9ex] \bottomrule 
	\end{tabular}

	\caption{The mean and $1\sigma$ of parameters for the $M_r<-21$ subsample. The different columns show fits down to different minimum scales including one with fixed Alcock--Paczynski parameters. 
	The table is vertically divided in four parts for ease of reading based on the nature of different parameters. The first set of rows showing base HOD parameters, the second set showing dynamical and satellite parameters, the third set for AP parameters to understand cosmology dependence, and the final set are some derived parameters.}
	\label{tab:HODMr21_wscale}
\end{table*}

\begin{table*}
	\centering
	\begin{tabular}{|c| c| c| c| c| c| c| c| c| c|} 
	\toprule
	  & \multicolumn{3}{|c|}{\vaststrut\bf free ($f_c,\gamma_{\rm IHV},f_{\rm vir}$)}  & \multicolumn{3}{|c|}{\bf free ($f_c,\gamma_{\rm IHV}$)}  & \multicolumn{3}{|c|}{\bf free ($f_{\rm vir}$)} \\ \hline
	 {\bf Parameters} & ${\bf M_r<-21}$  & ${\bf M_r<-20}$  & ${\bf M_r<-19}$  & ${\bf M_r<-21 }$ & ${\bf M_r<-20 }$ & ${\bf M_r<-19 }$ & ${\bf M_r<-21}$  & ${\bf M_r<-20}$  & ${\bf M_r<-19}$  \\ [0.8ex] 
	 \multicolumn{10}{|l|}{\bf Basic HOD model} \\ \hdashline
	$\log_{10}(M_{\rm cut})$ & $12.64^{+0.1}_{-0.07}$  & $12.1^{+0.13}_{-0.1}$  & $11.86^{+0.25}_{-0.22}$  & $12.69^{+0.08}_{-0.08}$  & $12.03^{+0.1}_{-0.08}$  & $11.72^{+0.19}_{-0.13}$  & $12.78^{+0.09}_{-0.08}$  & $12.25^{+0.12}_{-0.12}$  & $11.98^{+0.16}_{-0.18}$  \\ [0.9ex]
	$\sigma_{\rm M}$ & $0.77^{+0.29}_{-0.31}$  & $1.12^{+0.3}_{-0.28}$  & $1.58^{+0.5}_{-0.51}$  & $0.99^{+0.2}_{-0.21}$  & $0.98^{+0.23}_{-0.25}$  & $1.34^{+0.42}_{-0.31}$  & $1.18^{+0.18}_{-0.18}$  & $1.42^{+0.22}_{-0.25}$  & $1.78^{+0.31}_{-0.34}$  \\ [0.9ex]
	$\log_{10}(M_1)$ & $13.58^{+0.15}_{-0.17}$  & $13.05^{+0.13}_{-0.15}$  & $12.66^{+0.21}_{-0.23}$  & $13.71^{+0.09}_{-0.1}$  & $13.24^{+0.09}_{-0.08}$  & $12.95^{+0.09}_{-0.09}$  & $13.37^{+0.17}_{-0.27}$  & $12.77^{+0.19}_{-0.24}$  & $12.24^{+0.24}_{-0.16}$  \\ [0.9ex]
	$\alpha$ & $0.71^{+0.26}_{-0.27}$  & $0.91^{+0.14}_{-0.14}$  & $1^{+0.15}_{-0.14}$  & $0.9^{+0.21}_{-0.19}$  & $1.05^{+0.12}_{-0.11}$  & $1.11^{+0.08}_{-0.08}$  & $0.59^{+0.2}_{-0.24}$  & $0.78^{+0.12}_{-0.12}$  & $0.71^{+0.13}_{-0.1}$  \\ [0.9ex]
	$\kappa$ & $2.31^{+0.48}_{-0.78}$  & $2.53^{+0.3}_{-0.45}$  & $2.29^{+0.47}_{-0.68}$  & $2.21^{+0.48}_{-0.61}$  & $2.21^{+0.52}_{-0.67}$  & $2.24^{+0.49}_{-0.42}$  & $2.38^{+0.42}_{-0.54}$  & $2.33^{+0.34}_{-0.54}$  & $2.35^{+0.45}_{-0.73}$  \\ [0.9ex]
	 \multicolumn{10}{|l|}{\bf Dynamics and Satellite} \\ \hdashline
	$\gamma_{\rm HV}$ & $1.32^{+0.12}_{-0.11}$  & $1.01^{+0.11}_{-0.1}$  & $0.95^{+0.14}_{-0.12}$  & $1.27^{+0.13}_{-0.12}$  & $1.07^{+0.1}_{-0.1}$  & $1.08^{+0.1}_{-0.1}$  & $1.29^{+0.13}_{-0.14}$  & $0.77^{+0.1}_{-0.09}$  & $0.83^{+0.1}_{-0.07}$  \\ [0.9ex]
	$f_{\rm c}$ & $0.26^{+0.25}_{-0.15}$  & $0.55^{+0.32}_{-0.24}$  & $1.01^{+0.58}_{-0.4}$  & $0.28^{+0.26}_{-0.18}$  & $0.47^{+0.25}_{-0.2}$  & $0.81^{+0.38}_{-0.25}$  &  fixed   &  fixed   &  fixed   \\ [0.9ex]
	$\gamma_{\rm IHV}$ & $0.99^{+0.2}_{-0.19}$  & $0.8^{+0.14}_{-0.12}$  & $0.8^{+0.17}_{-0.15}$  & $0.84^{+0.16}_{-0.12}$  & $0.73^{+0.09}_{-0.09}$  & $0.68^{+0.1}_{-0.09}$  &  fixed   &  fixed   &  fixed   \\ [0.9ex]
	$f_{\rm vir}$ & $1.27^{+0.35}_{-0.27}$  & $1.56^{+0.46}_{-0.34}$  & $2.75^{+1.12}_{-0.76}$  &  fixed   &  fixed   &  fixed   & $1.81^{+0.6}_{-0.38}$  & $3.4^{+0.84}_{-0.78}$  & $3.91^{+0.74}_{-0.94}$  \\ [0.9ex]
	 \multicolumn{10}{|l|}{\bf Alcock--Paczynski parameters} \\ \hdashline
	$\alpha_{\parallel}$ & $1^{+0.03}_{-0.03}$  & $0.99^{+0.03}_{-0.03}$  & $1.03^{+0.02}_{-0.03}$  & $1.01^{+0.03}_{-0.03}$  & $1.01^{+0.02}_{-0.02}$  & $1.03^{+0.02}_{-0.02}$  &  fixed   &  fixed   &  fixed   \\ [0.9ex]
	$\alpha_{\perp}$ & $0.93^{+0.03}_{-0.03}$  & $0.99^{+0.02}_{-0.02}$  & $0.99^{+0.02}_{-0.02}$  & $0.93^{+0.03}_{-0.03}$  & $1^{+0.04}_{-0.03}$  & $1^{+0.02}_{-0.03}$  &  fixed   &  fixed   &  fixed   \\ [0.9ex]
	 \multicolumn{10}{|l|}{\bf Inferred Parameters} \\ \hdashline
	$f_{\rm sat}$ & $0.17 \pm 0.03$  & $0.19 \pm 0.03$  & $0.24 \pm 0.05$  & $0.12 \pm 0.01$  & $0.15 \pm 0.02$  & $0.15 \pm 0.02$  & $0.16 \pm 0.02$  & $0.23 \pm 0.02$  & $0.28 \pm 0.03$  \\ [0.9ex]
	$10^{3} \bar{n} \, [\mpcoh]^{-3}$ & $1.62$  & $6.73$  & $16.80$  & $1.61$  & $6.62$  & $16.20$  & $1.64$  & $6.95$  & $17.11$  \\ [0.9ex]
	$f\sigma_8(z_{\rm mean})$ & $0.61^{+0.06}_{-0.05}$  & $0.46^{+0.05}_{-0.05}$  & $0.43^{+0.06}_{-0.05}$  & $0.59^{+0.06}_{-0.06}$  & $0.49^{+0.05}_{-0.05}$  & $0.49^{+0.05}_{-0.05}$  & $0.60^{+0.06}_{-0.07}$  & $0.35^{+0.05}_{-0.04}$  & $0.37^{+0.05}_{-0.03}$  \\ [0.9ex] \bottomrule 
	\end{tabular}
	\strut\\
	\begin{tabular}{|c| c| c| c| c| c| c|} \toprule
	  & \multicolumn{3}{|c|}{\vaststrut {\bf free} ($f_c$) with: ${f_c \rightarrow  \gamma_{\rm IHV}}$ }  & \multicolumn{3}{|c|}{{\bf free} ($f_c$, ssp) with: $f_c \rightarrow  \gamma_{\rm IHV}$}  \\ \hline
	 {\bf Parameters} & ${\bf M_r<-21}$  & ${\bf M_r<-20}$  & ${\bf M_r<-19}$  & ${\bf M_r<-21}$  & ${\bf M_r<-20}$  & ${\bf M_r<-19}$  \\ [0.8ex] 
	 \multicolumn{7}{|l|}{\bf Basic HOD model} \\ \hdashline
	$\log_{10}(M_{\rm cut})$ & $12.64^{+0.08}_{-0.08}$  & $11.97^{+0.11}_{-0.07}$  & $11.65^{+0.12}_{-0.1}$  & $12.61^{+0.07}_{-0.06}$  & $12.03^{+0.12}_{-0.08}$  & $11.55^{+0.11}_{-0.09}$  \\ [0.9ex]
	$\sigma_{\rm M}$ & $0.84^{+0.21}_{-0.26}$  & $0.71^{+0.32}_{-0.27}$  & $1.13^{+0.27}_{-0.27}$  & $0.75^{+0.24}_{-0.24}$  & $0.92^{+0.3}_{-0.25}$  & $0.88^{+0.31}_{-0.34}$  \\ [0.9ex]
	$\log_{10}(M_1)$ & $13.7^{+0.1}_{-0.1}$  & $13.12^{+0.06}_{-0.06}$  & $12.73^{+0.07}_{-0.1}$  & $13.59^{+0.09}_{-0.08}$  & $13^{+0.07}_{-0.08}$  & $12.65^{+0.05}_{-0.06}$  \\ [0.9ex]
	$\alpha$ & $0.79^{+0.17}_{-0.18}$  & $0.86^{+0.08}_{-0.09}$  & $0.92^{+0.07}_{-0.08}$  & $0.62^{+0.14}_{-0.13}$  & $0.72^{+0.08}_{-0.09}$  & $0.85^{+0.05}_{-0.05}$  \\ [0.9ex]
	$\kappa$ & $2.12^{+0.58}_{-0.81}$  & $2.2^{+0.47}_{-0.55}$  & $2.56^{+0.3}_{-0.35}$  & $2.04^{+0.6}_{-1.16}$  & $1.84^{+0.62}_{-0.78}$  & $2.12^{+0.57}_{-0.82}$  \\ [0.9ex]
	 \multicolumn{7}{|l|}{\bf Dynamics and Satellite} \\ \hdashline
	$\gamma_{\rm HV}$ & $1.23^{+0.09}_{-0.1}$  & $1.12^{+0.08}_{-0.11}$  & $1.07^{+0.1}_{-0.08}$  & $1.22^{+0.09}_{-0.09}$  & $1.04^{+0.11}_{-0.1}$  & $1.07^{+0.09}_{-0.08}$  \\ [0.9ex]
	$f_{\rm c}$ & $0.12^{+0.2}_{-2.18}$  & $0.13^{+0.1}_{-1.02}$  & $0.33^{+0.12}_{-0.11}$  & $0.19^{+0.19}_{-0.53}$  & $0.21^{+0.09}_{-0.08}$  & $0.32^{+0.12}_{-0.1}$  \\ [0.9ex]
	$\gamma_{\rm IHV}$ &  fixed   &  fixed   &  fixed   &  fixed   &  fixed   &  fixed   \\ [0.9ex]
	$f_{\rm vir}$ &  fixed   &  fixed   &  fixed   &  fixed   &  fixed   &  fixed   \\ [0.9ex]
	 \multicolumn{7}{|l|}{\bf Alcock--Paczynski parameters} \\ \hdashline
	$\alpha_{\parallel}$ & $1.02^{+0.03}_{-0.03}$  & $1.03^{+0.02}_{-0.02}$  & $1.04^{+0.01}_{-0.01}$  & $1.01^{+0.04}_{-0.03}$  & $1.04^{+0.02}_{-0.01}$  & $1.04^{+0.01}_{-0.01}$  \\ [0.9ex]
	$\alpha_{\perp}$ & $0.92^{+0.03}_{-0.03}$  & $1.04^{+0.04}_{-0.06}$  & $1.02^{+0.02}_{-0.02}$  & $0.91^{+0.03}_{-0.03}$  & $1.04^{+0.03}_{-0.03}$  & $1.02^{+0.03}_{-0.02}$  \\ [0.9ex]
	 \multicolumn{7}{|l|}{\bf Inferred Parameters} \\ \hdashline
	$f_{\rm sat}$ & $0.14 \pm 0.01$  & $0.18 \pm 0.01$  & $0.2 \pm 0.02$  & $0.18 \pm 0.03$  & $0.21 \pm 0.02$  & $0.23 \pm 0.02$  \\ [0.9ex]
	$10^{3} \bar{n} \, [\mpcoh]^{-3}$ & $1.62$  & $6.53$  & $16.92$  & $1.72$  & $6.87$  & $17.80$  \\ [0.9ex]
	$f\sigma_8(z_{\rm mean})$ & $0.57^{+0.04}_{-0.05}$  & $0.51^{+0.04}_{-0.05}$  & $0.48^{+0.05}_{-0.04}$  & $0.57^{+0.04}_{-0.04}$  & $0.48^{+0.05}_{-0.05}$  & $0.48^{+0.04}_{-0.04}$  \\ [0.9ex] \bottomrule 
	\end{tabular}
 	\caption{The same as table~\ref{tab:HODMr21_wscale} but with different choice of extended HOD parameters. For each set of additional freedom we also allow all the base HOD parameters to be free. Each of the three absolute magnitude limited samples ($M_r<-21$,$M_r<-20$,$M_r<-19$) is shown for each set of free parameters.}
	\label{tab:HOD_wMr}
\end{table*}

\subsection{Galaxy host halo properties}
For each of the three limits in $r$ band absolute magnitude we fit for the occupation distribution in terms of $M_{\rm cut}$, which defines the $50\%$ probability of hosting a central galaxy, and $M_1$, which determines the characteristic halo mass for satellite galaxies. We obtain $\log_{10}(M_{\rm cut})= 11.86^{+0.25}_{-0.22},12.1^{+0.13}_{-0.1},12.64^{+0.1}_{-0.07} \msolaroh$ and $\log_{10}(M_{1})= 12.66^{+0.21}_{-0.23},13.05^{+0.13}_{-0.15},13.58^{+0.15}_{-0.17} \msolaroh$ for samples with absolute magnitude limit of $-19$, $-20$ and $-21$ respectively. We recover the well known correlation between host halo mass and galaxy luminosity. Our measurements are consistent with the results reported in \cite{guo2015b}, which used a similar approach to model redshift space galaxy clustering for SDSS data. We obtain satellite fractions of $f_{\rm sat}=0.24, 0.19 $ and $0.17$ for absolute magnitude limited samples at $-19$, $-20$ and $-21$ respectively. The satellite fraction we obtained is consistent with \cite{guo2015b}, but slightly smaller than the values reported in \cite{2011ApJ...736...59Z} and \cite{2013MNRAS.430..767C}. We note from the top left panel of Figure~\ref{fig:HODbase} that the characteristic halo mass of central galaxies ($M_{\rm cut}$) and satellite galaxies ($M_1$) are not correlated with each other. Also, the characteristic halo mass of satellites is strongly constrained using scales between $1\mpcoh$ and $5\mpcoh$ as shown by the difference between the black and silver contours. The positive correlation between $M_{\rm cut}$ and $\sigma_M$ as shown in the top middle panel of Figure~\ref{fig:HODbase} can be understood in terms of increasing $M_{\rm cut}$ increasing the galaxy bias and reducing the number density. Therefore to conserve both we need to increase $\sigma_M$, which allows more low mass haloes to host galaxies, resulting in a decrease in galaxy bias and increase in number density. Similarly, the positive correlation between $M_1$ and $\alpha$ as shown in the top right panel of Figure~\ref{fig:HODbase} can be understood in terms of increasing $M_1$ reducing the number of satellites, resulting in increased galaxy bias and decreasing fraction of satellite galaxies; both of these can be balanced by increasing $\alpha$. Our constraint on $\sigma_M$ is larger than previously reported values \citep[e.g.][]{2011ApJ...736...59Z,guo2015b}. This difference is probably caused by our need to fit very small scale clustering, which none of the previous studies incorporated. As one can see from table~\ref{tab:HODMr21_wscale} and the top middle panel of Figure~\ref{fig:HODbase}, the value of $\sigma_M$ increases as we include smaller scales in our analysis and our results on $\sigma_M$ using data above $1\mpcoh$ for the $M_r<-21$ subsample are consistent with the values reported in \cite{guo2015b}. This is probably because as we push clustering to the highly non-linear regime inside the halo, we become more sensitive to the presence of less massive haloes, which can be allowed for by increasing $\sigma_M$. This also leads to the interesting trend of a slight increase in $\sigma_M$ as we move from brighter galaxies ($M_r<-21$) to fainter ones ($M_r<-19$), which is opposite to the results seen in \cite{guo2015b}. We again think that this reflects the sensitivity to smaller haloes that arises from including clustering measurements to very small scales. We emphasise that our ability to model extremely small scales, together with the robustness of HOD parameters against choice of satellite galaxy degrees of freedom, shows that the simplified models regarding galaxy halo connection assumed in most large scale structure studies are valid, and that hydrodynamical processes seems not to affect the occupation probability of dark matter haloes. In later sections we will investigate whether baryonic physics has an impact on the more detailed kinematics and dynamics of galaxies beyond the occupation probability.

\subsection{Growth rate ($f\sigma_8$)}
We allow the velocity of galaxies to be scaled by a free parameter $\gamma_{\rm HV}$ which constrains the product of growth rate and amplitude of matter density fluctuation ($f\sigma_8$). The constraints on the growth rate of structure formation are reported in Table~\ref{tab:HODMr21_wscale} for the $M_r<-21$ subsample with different minimum scales and in Table~\ref{tab:HOD_wMr} for all three subsamples with various choices of satellite galaxy degrees of freedom. 
By allowing extra degree of freedom in the galaxy peculiar velocity, we constrain the deviation from GR in the growth rate of structure through $\gamma_{\rm HV}=0.95 \pm 0.13 (\rm stat.) \pm 0.02 (\rm sys.)$. This implies a constraint on the growth rate $f\sigma_8(z=0.2)=0.43 \pm 0.05$ under the assumption of $\Lambda$CDM and Planck cosmology for our deepest sample with the most flexible model. This is a $12\%$ constraint on growth rate in comparison to $25\% (14\%)$ obtained using multi-tracer analysis from large scales \citep{2013MNRAS.436.3089B} at $z=0.18(0.38)$. The constraint on the growth rate thus shows a modest improvement by pushing to very small scales, even though we lose precision by using only a single tracer. 
We show that this constraint on the growth rate is insensitive to choice in the freedom of satellite galaxy population or to fiducial cosmology via the Alcock--Paczynski parameters. The only sensitivity we find is that the brightest sample shows about a $2\sigma$ higher value of the growth rate. The brightest sample could be more sensitive to the non-linear effects and hence this could be an indication of assembly bias or aspect of galaxy or galaxy cluster formation not captured in our HOD model. Our constraint on the growth rate is also uncorrelated with the Alcock--Paczynski parameters, unlike large-scale or high redshift measurements; allowing the AP parameters to be free neither weakens our constraint nor introduces any bias. The measured $f\sigma_8$ is consistent with the prediction from Planck under the $\Lambda$CDM assumption with GR. The robustness of the growth rate measurement shows the great promise of the present approach in pushing cosmological constraints to very high precision in future studies. 

We have not explored all aspects of the satellite galaxy phase space distribution. It will be interesting to explore aspects discussed earlier that we chose to ignore in the present study owing to their relatively small impact, which will be very hard to see in the GAMA data. One can try to improve our measurement by combining results from GAMA and the SDSS Main Galaxy Sample (MGS) which covers a much larger volume but suffers from incompleteness at small scales. In the near future, the DESI BGS sample will be an ideal case to extend this analysis but will require more work including testing the model to much higher precision and employing simulations
of much larger volumes.

\subsection{Satellite galaxy properties}
 Apart from basic HOD parameters we are particularly interested in the details of the distribution and dynamics of satellite galaxies inside the dark matter haloes. These aspects of the model are studied using three additional parameters $f_c, \gamma_{\rm IHV}, f_{\rm vir}$ and two additional parameters for Alcock--Paczynski effect $\alpha_{\parallel}, \alpha_{\perp}$ shown in Figure \ref{fig:HODextended}. First we note that we obtain significant constraints on the AP parameters $\alpha_{\parallel}=1.0 \pm 0.02, \alpha_{\perp}=0.95 \pm 0.03$, consistent with the $\Lambda$CDM Planck predictions.
 
\subsubsection{Satellite concentration and velocity dispersion}

The satellite galaxies are distributed with the concentration of 1/10th for the $M_r<-21$ sample which increases to 1/3rd for the $M_r<-19$ sample compared to dark matter halo concentrations in the absence of any other satellite degree of freedom and requiring dynamical equilibrium to set $\gamma_{\rm IHV}$. This is probably unrealistic as it ignores all the physics especially to do with baryonic processes. Therefore, we also allow the velocity dispersion of satellite galaxies to be free on top of the concentration, and do not invoke the relationship between them for dynamical equilibrium with the dark matter distribution. The top left panel in Figure \ref{fig:HODextended} shows these two parameters and rules out the satellites following the dark matter distribution at high significance. Note that the satellite concentration is parameterized with parameter $f_c$ and detailed constraints are given in table~\ref{tab:HOD_wMr}. But this has an assumption that the satellite galaxies are limited within the virial radius of the dark matter haloes, which may not be the case. If we do allow the satellites to reside beyond the virial radius (by introducing $f_{\rm vir}$) then we expand the contours in $f_c-\gamma_{\rm IHV}$ space to include the possibility of satellite being populated following the dark matter distribution. In the absence of such an additional degree of freedom, we note that the brighter galaxies prefer to populate haloes with low concentration but consistent velocity dispersion, whereas the fainter satellite galaxies prefer to redistribute following the halo concentration but with a smaller velocity dispersion.

\subsubsection{Satellite boundary}
The analytic NFW formula for the density profiles of dark matter haloes lacks a sharp boundary and extends to infinity. But for dynamical reasons the virial radius seems like a sensible length to use as the definition of the halo boundary. Most HOD studies assume that the satellite galaxies in a dark matter haloes are populated no further than the virial radius. This is not necessarily true, and the issue will become increasingly important with more precise data. Therefore, we introduced the free parameter $f_{\rm vir}$ which defines the halo boundary as $f_{\rm vir}r_{\rm vir}$. This allows us to look at its impact on small scale clustering and whether this has a degeneracy with other satellite properties, such as concentration and velocity dispersion. The open contours in Figure \ref{fig:HODextended} represent the constraint while including this free parameter. We notice from the top left panel that allowing this freedom moves the contours slightly such that satellite galaxies follow the dark matter concentration and velocity dispersion within $2\sigma$. We find the halo boundary for the purpose of satellite galaxies depends on the limiting absolute magnitude of satellite galaxies. The brightest satellite galaxies ($M_r<-21$) are limited within the virial radius whereas fainter satellite galaxies ($M_r<-19$) extends to twice the virial radius. This result shows that satellite galaxies with different brightness/mass could be distributed at different distances from the halo centre. Note that in general including fainter galaxies makes the concentration and velocity dispersion consistent with dark matter haloes, but extends them to larger distances than the virial radius.

\subsubsection{Satellite statistics}
The number of satellite galaxies for a given halo depends on its dark matter mass in our HOD model given by equation \ref{eqn:HOD}. But this gives the mean occupation of the number of satellite galaxies and we still need to assume a statistical distribution, which is Poisson in our fiducial model. In principle given that galaxy formation correlates various scales with each other, the number statistics could deviate from Poisson and this is an important aspect to test. Therefore, we introduce an additional parameter, $A_p$, which allows the statistics of satellites to change between sub-Poisson ($A_p<1$) to super-Poisson ($A_p>1$) in a continuous manner. We allow this additional parameter to be free for each of the absolute magnitude limited samples, finding that the data do not constraint $A_p$ independent of the absolute magnitude limit. For the purpose of current observation we thus do not have the ability to constraint the satellite statistics. The last set of results in table \ref{tab:HOD_wMr} presents the constraints on all the parameters when $A_p$ is allowed to vary. The details of how the modified poission distributions were generated is given in Appendix~\ref{sec:modPois}.

\section{Summary and discussion}
\label{sec:summary}

We have explored a flexible approach that is designed to model the redshift-space clustering of galaxies in a manner that permits the inclusion of many realistic complications. The interest in RSD stems primarily from the original realisation \citep{Kaiser87} that the distortions in the linear regime had a simple analytic form that allowed the extraction of the logarithmic growth rate of density fluctuations (or at least the related parameter $f\sigma_8$). But subsequent work has served to emphasise that nonlinear and quasilinear modifications of the Kaiser result exist up to rather large scales, so that precise models for the RSD effect down to small scales are required for efficient statistical exploitation of RSD measurements from galaxy redshift surveys. Ideally, such models would be analytic in nature, and much heroic effort has been invested in extending perturbation theory in this direction (see e.g. \citejap{2017JCAP...10..009H}). Our concern, however, has been that it is difficult in such work to include some of the more complicated aspects of the relation between galaxies and the underlying dark matter. We have therefore taken a brute-force approach in which model galaxies are inserted into simulated dark-matter distributions according to various recipes. These recipes are governed by a set of nuisance parameters, and we have three related aims: (1) how well can such degrees of freedom be constrained by data? (2) how large is the potential bias in the growth rate if these complications are neglected? (3) how does the precision in the recovered growth rate degrade if we marginalise over all the hidden degrees of freedom? 

Our approach has been to work with an extension of the Halo Model, using empirical catalogues of dark-matter haloes extracted from the Bolshoi and MultiDark Planck (MDPL1)  
simulations \citep{2011ApJ...740..102K,2012MNRAS.423.3018P}. A simple model would define a Halo Occupation Distribution function, $N(M)$, which specifies the existence (or not) of a central galaxy and a number of satellites as a function of halo mass. Such a model has met with a good deal of success in accounting for low-order galaxy correlations, but it makes many unwarranted assumptions. The ones that we have focused on in this work are spatial and kinematical biases: should we assume that the satellites are distributed within a halo in the same way as the dark matter, and do their peculiar velocities have the same
distribution function as the dark matter? To the extent that haloes are equilibrium objects, any such biases should be coupled: satellites that are `cooler' than the dark matter will sink to the centre of the halo, and vice versa. We have therefore introduced nuisance parameters that explicitly govern such biases. We have also allowed a large-scale velocity bias, in which all peculiar velocities arising from the centre-of-mass motion of haloes are scaled. This is not expected to be an astrophysical degree of freedom in the same way as the small-scale velocity bias, but rather would reflect a large-scale modification of the strength of gravity. The whole interest in measuring RSD arises from the desire to test such modifications, so it is important to be able to generate mock galaxy datasets in which they are present.

We have applied this modelling to data from the GAMA survey (\citejap{2015MNRAS.452.2087L}; \citejap{2018MNRAS.474.3875B}), which is a flux limited spectroscopic survey of approximately 300,000 galaxies over $230$ deg$^2$. Larger catalogues of greater statistical power exist, but for our present purposes the GAMA dataset has the unique advantage of being close to 100\% in spectroscopically complete. Thus the small-scale clustering measurements are not compromised by the common problem of missing close pairs, and we can see how well our modelling works down to the smallest scales.

We first present the constraint on the distribution of galaxies' host halo mass and their brightness. We find that such distributions are not sensitive to clustering data below $1\mpcoh$. We also show that these distributions are independent of what goes on within a halo in detail and hence the occupation probability of dark matter haloes with galaxy brightness is insensitive to the baryonic physics. We then present our constraint on the growth rate $f\sigma_8(z=0.2)=0.43 \pm 0.05$, which is more precise than previous constraints from same dataset \citep{2013MNRAS.436.3089B}. We note that the constraint on the growth rate is insensitive to whether or not we allow the additional degrees of freedom regarding the detailed kinematics of the galaxies within dark matter haloes. We finally present one of the very first constraints on the detailed kinematic degree of freedom for galaxies within haloes. We found that brighter galaxy samples show significantly lower satellite concentrations, but have NFW-level velocity dispersions; in contrast, fainter samples show NFW-level satellite concentrations but have lower velocity dispersions. This deviation of satellite kinematics from the properties of the underlying NFW dark matter halo reduces if we allow the satellite boundary to be beyond the virial radius. We then find that the satellite concentration and velocity dispersion become consistent with NFW kinematics at the $2\sigma$ level, but there is then a trend in the boundary with the depth of the sample -- with the satellite distribution for the deeper sample tending to extend to larger radii.

It is encouraging that our results provide the first constraints on the kinematics of satellites from a cosmologically significant volume. We found no significant potential bias in the measurement of the growth rate, nor degradation in error while marginalising over the hidden parameters of galaxy formation. Our main limitation here is the use of a conservative covariance matrix, which degrades our constraints, and also the limited volume covered by GAMA. A more sophisticated treatment of the covariance matrix will be possible in the future as more high resolution N-body simulations become available. In terms of progress on data, the Bright Galaxy Survey as part of DESI \citep{2016arXiv161100036D} will have a very similar flux selection to GAMA, but will cover 14000 square degrees -- thus approximately 100 times the survey volume.  But unlike GAMA, DESI will be strongly incomplete for close pairs, with a major impact on its measurements of small scale clustering. Extraction of the RSD signal from DESI will therefore require either a sacrifice of precision through restricting the analysis to very large scales, or a reliance on sophisticated schemes for cancelling the  incompleteness (see \citejap{Smith2019}). In either case, however, it will be important to be able to confront the results with mock data that include real-world effects of galaxy formation. We feel that our extended treatment of satellite properties in the Halo Model framework will be able to make a valuable contribution here.

Finally, an interesting extension of this work will be to use additional statistics measured from the same dataset as complementary information in the RSD analysis. One such statistic recently proposed in \cite{2020arXiv200108760P} is the Voronoi Volume Function (VVF), which is in effect a convenient combination of higher-order correlations. It will be interesting to see if one can improve the constraints presented in this paper by including the VVF or other similar  statistics. We will leave such possibilities for future work.

\section{Data Availability}
All of the observational datasets used in this paper will be made available through the GAMA website \url{http://www.gama-survey.org/}. The codes used in this analysis along with instructions will be made available on \url{https://www.roe.ac.uk/~salam/CodeData/} on publication. The N-body simulations used in this paper can be accessed through \url{https://www.cosmosim.org/cms/simulations/bolshoi/}.

\section*{Acknowledgments}

We thank Aseem Paranjape for insightful discussions. We thank Horst Meyerdierks and Eric Tittley for their support with Stacpolly and Cuillin cluster where all of the computing for this project is performed. 
SA and JAP are supported by the European Research Council
through the COSFORM Research Grant (\#670193).  We thank
the Multi Dark Patchy Team for making their simulations publicly available. This research has made use of NASA's Astrophysics Data System. SA was supported in part by the International Centre for Theoretical Sciences (ICTS) during a visit for participating in the programme `Cosmology -- The Next Decade' (Code: ICTS/cosmo2019/01).

GAMA is a joint European-Australasian project based around a spectroscopic campaign using the Anglo-Australian Telescope. The GAMA input catalogue is based on data taken from the Sloan Digital Sky Survey and the UKIRT Infrared Deep Sky Survey. Complementary imaging of the GAMA regions is being obtained by a number of independent survey programmes including GALEX MIS, VST KiDS, VISTA VIKING, WISE, Herschel-ATLAS, GMRT and ASKAP providing UV to radio coverage. GAMA is funded by the STFC (UK), the ARC (Australia), the AAO, and the participating institutions. The GAMA website is \url{http://www.gama-survey.org/} . 

The CosmoSim database used in this paper is a service by the Leibniz-Institute for Astrophysics Potsdam (AIP).
The MultiDark database was developed in cooperation with the Spanish MultiDark Consolider Project CSD2009-00064.

The authors gratefully acknowledge the Gauss Centre for Supercomputing e.V. (\url{www.gauss-centre.eu}) and the Partnership for Advanced Supercomputing in Europe (PRACE, \url{www.prace-ri.eu}) for funding the MultiDark simulation project by providing computing time on the GCS Supercomputer SuperMUC at Leibniz Supercomputing Centre (LRZ: \url{www.lrz.de}).


\bibliography{Master_Shadab}
\bibliographystyle{mnras}
\appendix

\section{Modified Poisson random numbers}
\label{sec:modPois}
\begin{figure}
    \centering
    \includegraphics[width=0.49\textwidth]{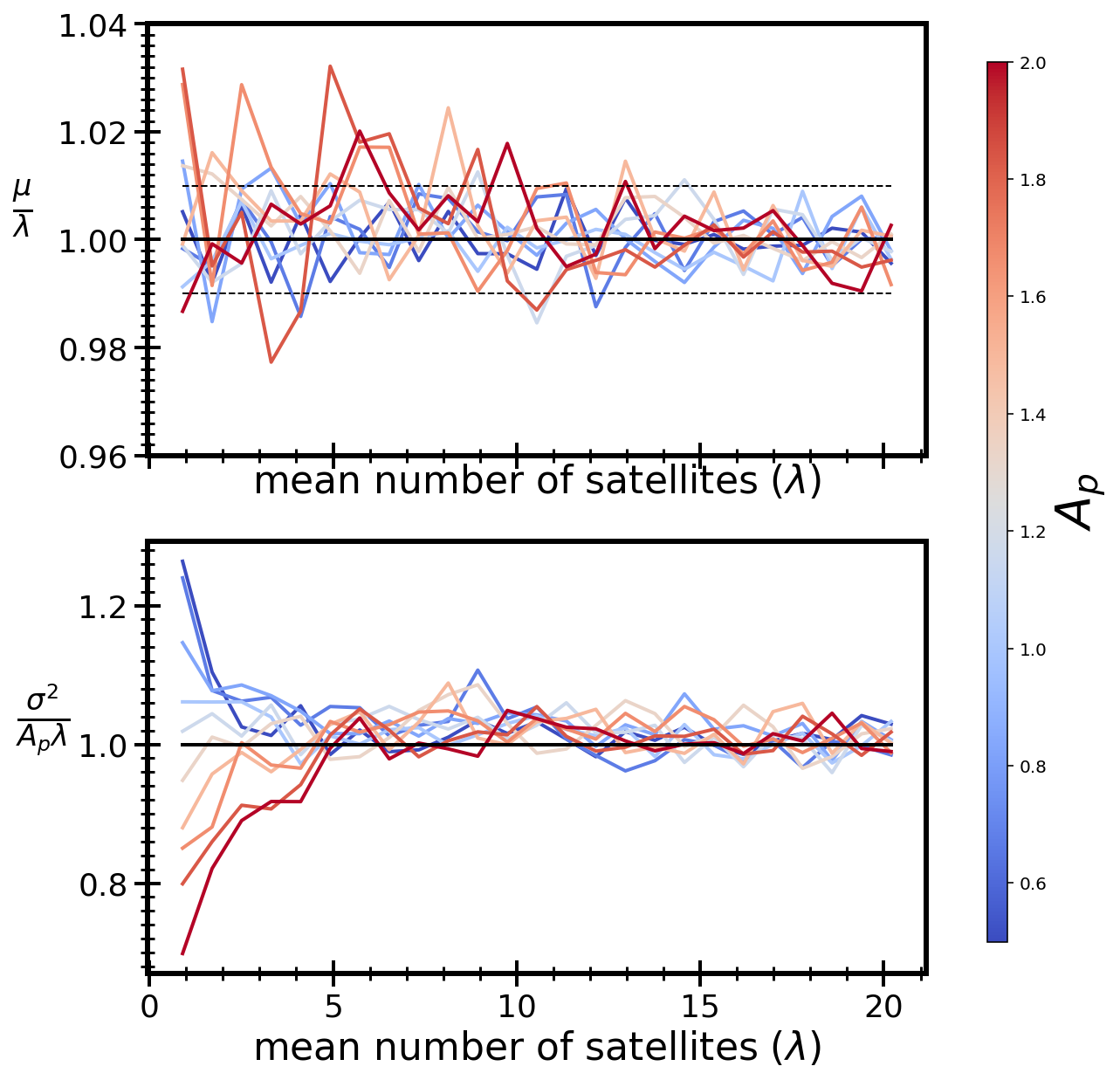}
    \caption{Accuracy of the modified Poisson distribution python library. The top panel shows the ratio of the mean of the random numbers to the expected mean as the function of parameter $\lambda$. The bottom panel shows the ratio of observed variance to the expected variance. The different coloured line shows the different values of the $A_p$ parameter.}
    \label{fig:poisson}
\end{figure}

We developed a python library to efficiently generate super- and sub-Poisson random numbers to study the nature of satellite galaxy statistics.
The Poisson distribution is given by following equation:
\begin{equation}
    f(k;\lambda)= \frac{\lambda^k e^{-\lambda}}{k!}
\end{equation}
where $k$ is the random integer with mean and variance $\lambda$. The goal is to generalise this to be able to generate random numbers with a similar distribution that preserves the same mean, but which has a different variance. In other words we want to obtain a distribution such that:
\begin{align}
    \mu &= \int k P(k; \lambda, A_p) dk = \lambda \\
    \sigma^2 &= \int k^2 P(k; \lambda, A_p) dk -\mu^2 = A_p \lambda 
\end{align}
where $A_p$ is an additional parameter controlling the width of the distribution. Therefore, $A_p=1$ will give a Poisson distribution, $A_p>1$ will lead to super-Poisson fluctuations, and $A_p<1$ will result in sub-Poisson fluctuations. We achieve this by first generating a Poisson distribution with $\lambda'=\lambda/A_p$ and then we map $k\rightarrow A_p \times k$. In this way we generate a distribution with variance $A_p\lambda$ and mean of $\lambda$. But the problem is that the random numbers generated in this way do not sample the distribution at integer values as $A_p$ is not an integer in general. Therefore, we interpolate this probability distribution to integer values to generate an integer sampled probability distribution satisfying our requirement. Figure~\ref{fig:poisson} shows the accuracy of this method with the top panel showing that the ratio of the mean of the distribution and the target mean is unity within 2\%. The bottom panel shows the variance is scaled by $A_p$ as per our requirement for $\lambda>4$ but it shows some deviation for smaller $\lambda$. The python library is available at \url{https://gitlab.com/shadaba/modified_poisson/-/blob/master/Example_usage_plot.ipynb}.

\label{lastpage}

\end{document}